\documentclass[10pt,twoside,twocolumn,journal]{IEEEtran}

\usepackage[utf8]{inputenc} 
\usepackage[T1]{fontenc}
\usepackage{url}
\usepackage{ifthen}
\usepackage{graphicx}
\usepackage[T1]{fontenc}

\usepackage{enumitem}
\usepackage[cmex10]{amsmath}
\usepackage{amssymb}

\usepackage{cite}
\usepackage{amsthm}
\usepackage{textcomp}
\usepackage{siunitx}
\usepackage{color}
\usepackage{cases}
\usepackage[font={small}]{caption}
\usepackage{epstopdf}
\usepackage{graphicx}
\usepackage{caption}
\usepackage{verbatim} 
\usepackage{bbm}
\usepackage{mathtools}

\usepackage[lofdepth,lotdepth]{subfig}
\usepackage[countmax]{subfloat}
\usepackage{tcolorbox}
\usepackage{lipsum}
\usepackage{etoolbox}
\AtBeginEnvironment{tcolorbox}{\small}
\usepackage{fix-cm}
\usepackage{bbm}
\usepackage{tcolorbox}
\usepackage{multirow}
\usepackage{tikz}
\usetikzlibrary{positioning}
\usepackage{booktabs} 

\graphicspath{{figure/}{figures/}}

\newtheorem{theorem}{Theorem}

\theoremstyle{definition}

\theoremstyle{remark}
\newtheorem{remark}{Remark}

\setlist[description]{style=multiline}

\begin{document}
\sloppy

\title{PolyShard: Coded Sharding Achieves Linearly Scaling Efficiency and Security Simultaneously}

\author{Songze Li, Mingchao Yu, Chien-Sheng Yang, A. Salman Avestimehr, Sreeram Kannan and Pramod Viswanath
\thanks{This material is based upon work supported by the Distributed Technology Research Foundation, Input-Output Hong Kong, the National Science Foundation under grants CCF-1705007, CCF-1763673 and CCF-1703575, and the Army Research Office under grant W911NF1810332. This material is also based upon work supported by Defense Advanced Research Projects Agency (DARPA)
under Contract No. HR001117C0053. The views, opinions, and/or findings expressed are those of the authors and should not
be interpreted as representing the official views or policies of the Department of Defense or the U.S. Government. A part of this paper was submitted to IEEE ISIT 2020. \textit{ (Corresponding author: Chien-Sheng Yang.)}}
\thanks{S.~Li, M.~Yu, C.-S.~Yang and A.~S.~Avestimehr are with the Department of Electrical and Computer Engineering, University of Southern California, Los Angeles, CA 90089 USA (e-mail: songzeli@usc.edu; mingchay@usc.edu; chienshy@usc.edu; avestimehr@ee.usc.edu).}%
\thanks{S.~Kannan is with the Department of Electrical and Computer Engineering Department, University of Washington, Seattle, WA 98195, USA (e-mail: ksreeram@uw.edu).}%
\thanks{P.~Viswanath is with the Department of Electrical and Computer Engineering Department, University of Illinois at Urbana-Champaign, IL 61820, USA (e-mail: pramodv@illinois.edu).}}

\maketitle

\begin{abstract}
Today's blockchain designs suffer from a trilemma claiming that no blockchain system can simultaneously achieve decentralization, security, and performance scalability. For current blockchain systems, as more nodes join the network, the efficiency of the system (computation, communication, and storage) stays constant at best. A leading idea for enabling blockchains to scale efficiency is the notion of sharding: different subsets of nodes handle different portions of the blockchain, thereby reducing the load for each individual node. However, existing sharding proposals achieve efficiency scaling by compromising on trust - corrupting the nodes in a given shard will lead to the permanent loss of the corresponding portion of data. In this paper, we settle the trilemma by demonstrating a new protocol for coded storage and computation in blockchains. In particular, we propose  {\tt PolyShard}: ``polynomially coded sharding'' scheme that achieves information-theoretic upper bounds on the efficiency of the storage, system throughput, as well as on trust, thus enabling a truly scalable system. We provide simulation results that numerically demonstrate the performance improvement over state of the arts, and the scalability of the {\tt PolyShard} system. Finally, we discuss potential enhancements, and highlight practical considerations in building such a system.
\end{abstract}
\section{Introduction}\label{sec:intro}
\begin{figure}[htbp]
\centering
    \includegraphics[width=0.8\linewidth]{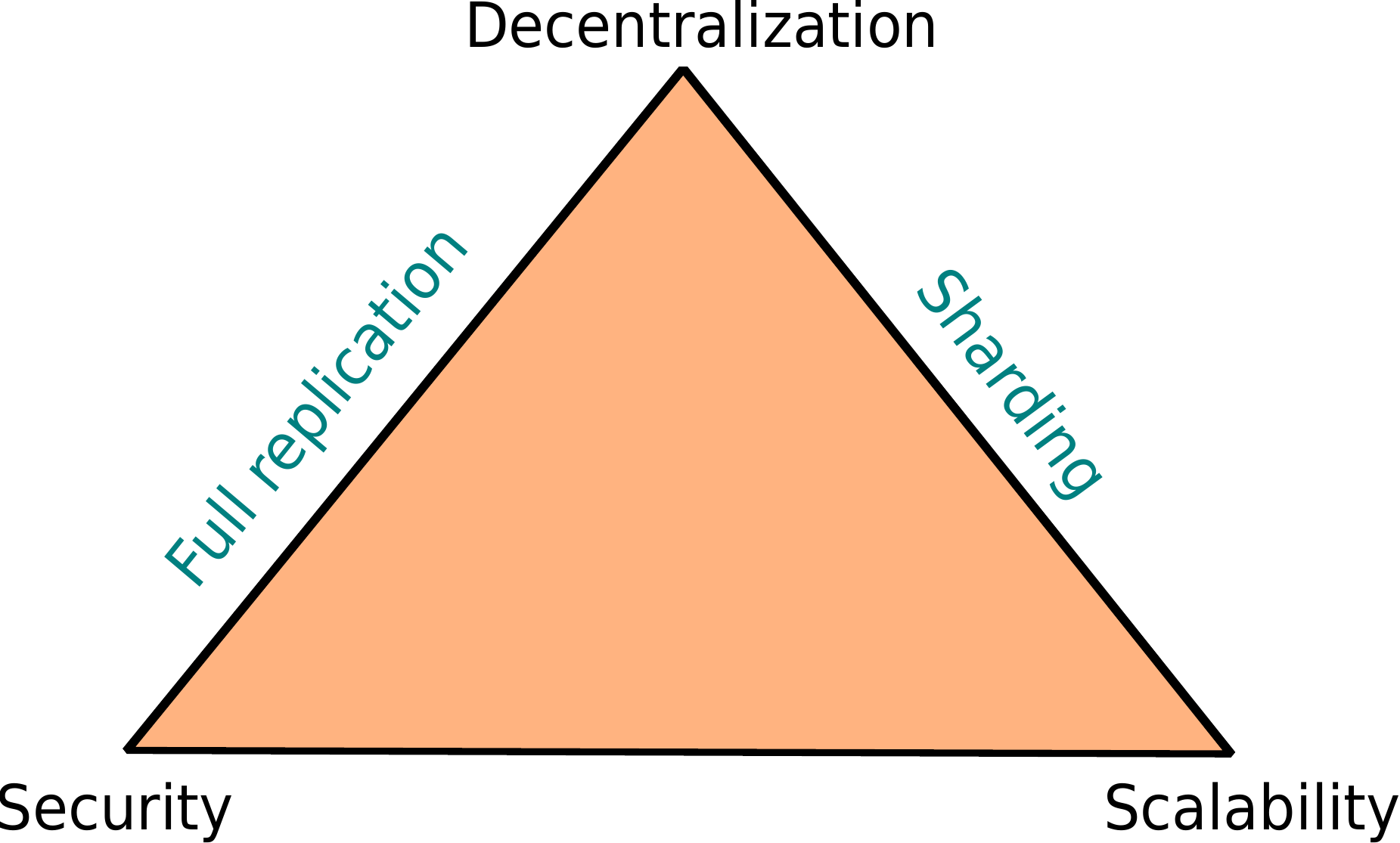}
    \caption{Blockchain trilemma. No current blockchain system can simultaneously achieve decentralization, security, and scalability.}
    \label{fig:trilemma}
\end{figure}

While Blockchain systems promise a host of new and exciting applications, such as digital cryptocurrency~\cite{nakamoto2008bitcoin}, industrial IoT~\cite{bahga2016blockchain}, and healthcare management~\cite{mettler2016blockchain}, their scalability remains a critical challenge \cite{croman2016scaling}. In fact, a well-known blockchain trilemma (see Figure~\ref{fig:trilemma}) has been raised~\cite{trilemma} claiming that no  decentralized ledger system can simultaneously achieve 1) security (against adversarial attacks), 2) decentralization (of computation and storage resources), and 3) scalability (of throughput with the network size). All existing major blockchains either achieve decentralization at the cost of efficiency, or efficiency at the cost of decentralization and/or security.

The focus of this paper is to formalize and study a version of the blockchain trilemma, in order to understand whether it is fundamental to blockchain systems. Corresponding to the three traits in the trilemma, we study the following performances measures of blockchain systems: security - measured as the number of malicious nodes the system can tolerate,   decentralization - measured as the fraction of the block chain (or ledger) that is stored and processed by each node (we denote its inverse by {\em storage efficiency}), and scalability - measured as the total \emph{throughput} of the system that is the number of computations performed in the system (e.g., number of transactions verified) in a unit period of time.

\begin{figure}[t]
\centering
    \subfloat[Full replication. 3 transactions verified per epoch. Can tolerate $\frac{N}{2}-1=14$ malicious nodes.]{\includegraphics[width=0.43\linewidth]{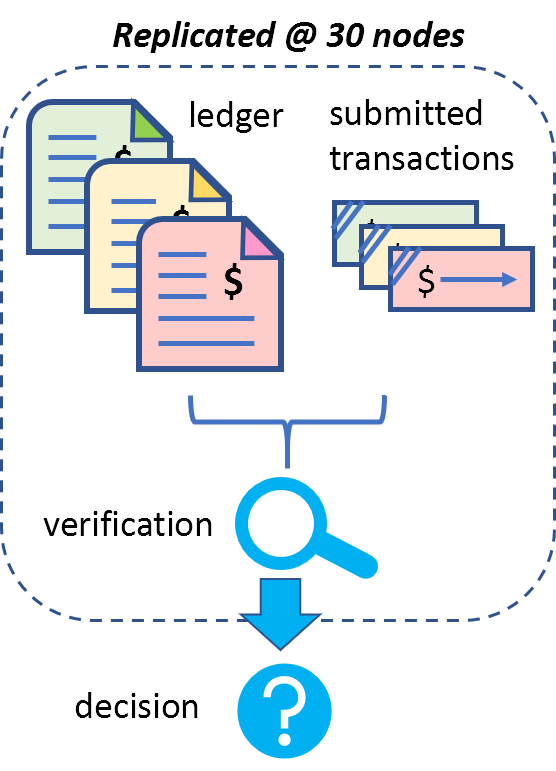}}\hspace{10pt}
    \subfloat[Sharding with 3 shards. 9 transactions verified per epoch. Can tolerate $\frac{N}{2K}-1=4$ malicious nodes.]{\includegraphics[width=0.5\linewidth]{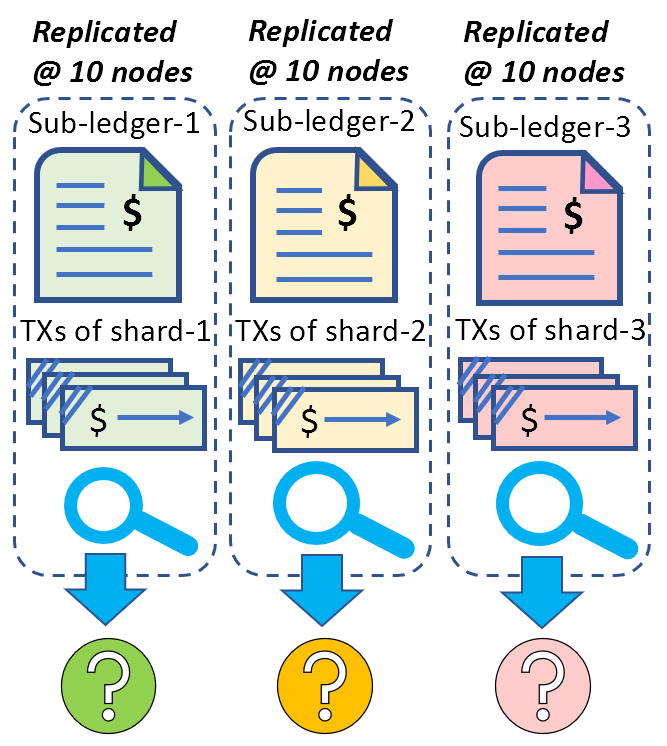}}
    \caption{A blockchain system with 30 nodes, each of which is capable of verifying 3 transactions per epoch. With 3-sharding, the ledger is partitioned into 3 sub-ledgers, and transactions are limited to between the accounts within the same sub-ledger. Sharding improves the storage and throughput efficiency by 3 time, at the cost of compromising security by about 3 times.}
    \label{fig:fr_and_sharding}
\end{figure}

Within this context, let us first examine the current blockchain systems. 
Bitcoin~\cite{nakamoto2008bitcoin} and Ethereum~\cite{wood2014ethereum} are designed based on a {\em  full replication} system, in which each network node  stores the entire blockchain and replicates all the computations (e.g., transaction verifications), as demonstrated in Figure \ref{fig:fr_and_sharding}(a). This feature enables  high security (of tolerating 49\% adversarial nodes), however drastically limits the storage efficiency and throughput of the system: they stay constant regardless of the number of nodes $N$. For example, Bitcoin currently restricts its block size to 1 MB, and processing rate to 7 transactions/sec~\cite{gervais2016security}. In practice, the computational burden even increases with $N$ (e.g., mining puzzles get harder as time progresses and more users participate), causing the throughput to drop. 

To scale out throughput and storage efficiency, the leading solution being discussed in the blockchain literature is via {\em sharding}  \cite{luu2016secure,kokoris2017omniledger,gencer2017short}. The key idea is to partition the blockchain into $K$ independent sub-chains, which are then replicated separately at $q=N/K$ nodes to yield $K$ smaller full-replication systems, a.k.a., shards (Figure \ref{fig:fr_and_sharding}(b)). This way, both storage efficiency and throughput are improved by a factor of $K$. However, to scale this improvement with network size, $K$ must increase linearly with $N$. Consequently, the number of nodes $q$ per shard has to be held constant, which allows an attacker to corrupt as few as $q/2$ nodes to compromise a shard. This yields a security level of $q/2$, which approaches zero as $N$ grows. Although various efforts have been made to alleviate this security issue (e.g., by periodically shuffling the nodes \cite{kokoris2017omniledger}), they are susceptible to powerful adversaries (e.g., who can corrupt the nodes {\em after} the shuffling), yet none scale security.



In summary, both full replication and sharding based blockchain systems make trade-offs between the scalability of throughput, storage efficiency, and security. However, such a trade-off is far from optimal from information theoretical point of view. Given the $N\times$ computation and $N\times$ storage resources across all the $N$ network nodes, the following information-theoretic upper bounds hold: $\textup{security} \leq \Theta(N)$; $\textup{throughput} \leq \Theta(N)$; $\textup{storage efficiency} \leq \Theta(N)$.
It is intuitive that these bounds can be \emph{simultaneously} achieved by a centralized system, allowing all the three metrics to scale. However, as pointed out by the trilemma, this has not been achieved by any existing decentralized system. This raises the following fundamental open problem: 

\begin{tcolorbox}
Is there a blockchain design that \emph{simultaneously} scales storage efficiency, security, and throughput?
\end{tcolorbox}

We answer this question affirmatively by introducing the concept of {\em coded sharding}. In particular, we propose  {\tt PolyShard} (polynomially coded sharding), a scheme that simultaneously  achieves linear scaling in throughput, storage efficiency, and security (i.e., $\Theta(N)$). We show mathematically that {\tt Polyshard} achieves all three information-theoretic upper bounds and enables a truly scalable blockchain system (see Table~\ref{table:compare}).
\begin{table}[h]
  \centering
  \scalebox{0.9}{
  \begin{tabular}{| c | c | c | c | }
    \hline
    \rule{0pt}{10pt} & Storage efficiency & Security & Throughput \\ \hline
    \rule{0pt}{10pt} Full replication & $O(1)$  & $\Theta(N)$  &  $O(1)$ \\ \hline
    \rule{0pt}{10pt}  Sharding & $\Theta(N)$ & $O(1)$  & $\Theta(N)$ \\ \hline
     \rule{0pt}{10pt} Information-theoretic limit & $\Theta(N)$  & $\Theta(N)$  & $\Theta(N)$ \\ \hline
     \rule{0pt}{10pt} {\tt PolyShard} (this paper) & $\Theta(N)$  & $\Theta(N)$  & $\Theta(N)$ \\ \hline
  \end{tabular}
  }
\caption{Performance comparison of the proposed {\tt PolyShard} verification scheme with other benchmarks and the information-theoretic limits.}
\label{table:compare}
\end{table}

\begin{figure}[t]
\centering
    \includegraphics[width=\linewidth]{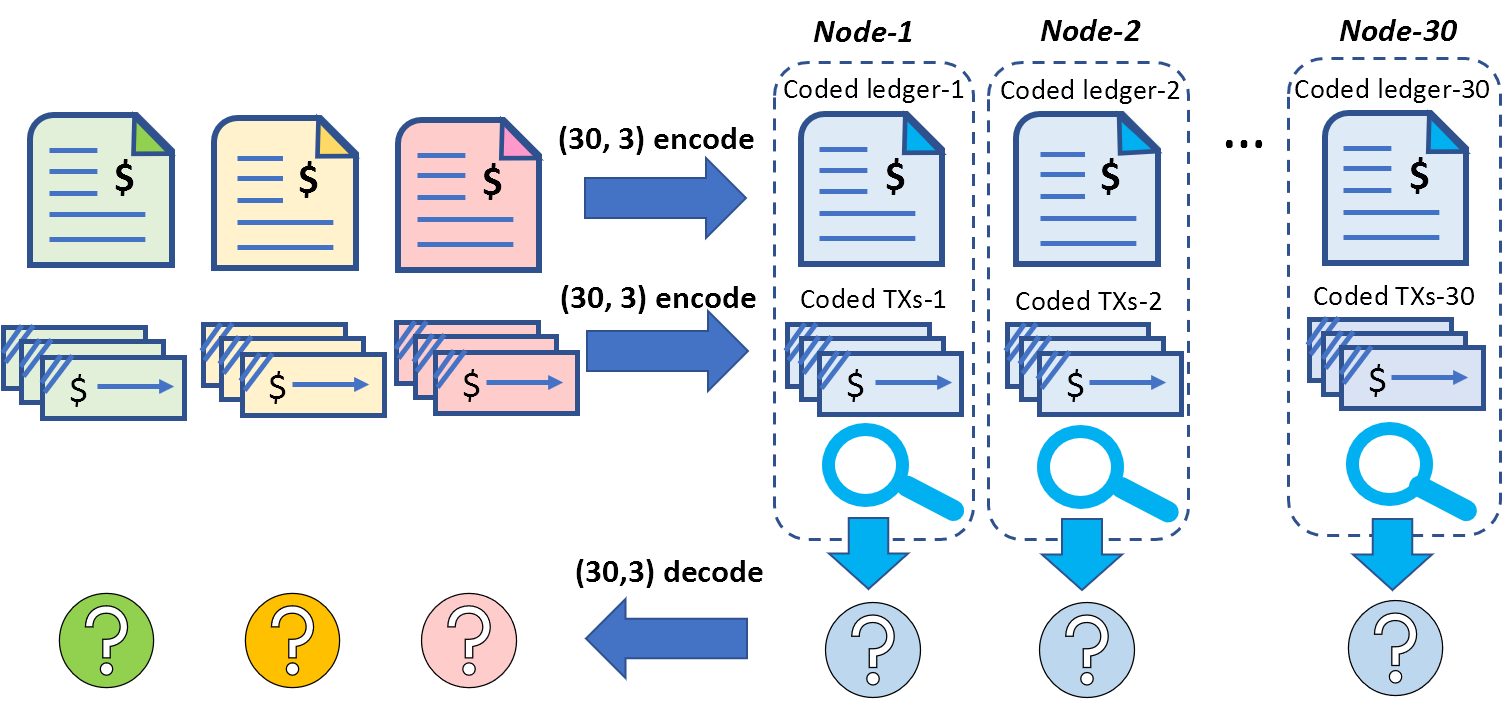}
    \caption{The proposed {\tt PolyShard} system with 30 nodes. Each node computes a verification function on a coded sub-ledger and a coded block, which are created by a distinct linear combination of the original sub-ledgers, and the proposed blocks respectively. Since encoding does not change the size of the sub-ledger, {\tt PolyShard} achieves the same storage and throughput efficiency (i.e., 9 transactions per epoch) as the conventional sharding solution. Additionally, {\tt PolyShard} improves the security guarantee to protect against $(N-K)/2=13$ malicious nodes, for degree-$1$ verification functions.}
    \label{fig:polyshard}
\end{figure}

{\tt PolyShard} is inspired by recent developments in \emph{coded computing}~\cite{LMA_all,li2016fundamental,lee2018speeding,LMA16_unify,dutta2016short,yu2017polynomial,karakus2017straggler,TLDK-ICML,yu2018lagrangeNIPS}, in particular Lagrange Coded Computing~\cite{yu2018lagrangeNIPS}, which provides a transformative framework for injecting computation redundancy in unorthodox coded forms in order to deal with failures and errors in distributed computing. The key idea behind {\tt PolyShard} is that instead of storing and processing a single uncoded shard as done conventionally, each node stores and computes on a coded shard of the same size that is generated by linearly mixing uncoded shards (Figure \ref{fig:polyshard}), using the well-known Lagrange polynomial. This coding provides computation redundancy to simultaneously provide security against erroneous results from malicious nodes, which is enabled by noisy polynomial interpolation techniques (e.g., Reed-Solomon decoding).



While  coding  is generally applicable in many distributed computing scenarios, the following two salient features make {\tt PolyShard} particularly suitable for blockchain systems.
\begin{itemize}[leftmargin=*]
    \item {\em Oblivious}: The coding strategy applied to generate coded shards is oblivious of the verification function. That means, the same coded data can be simultaneously used for multiple verification items (examples: digital signature verification and balance checking in a payment system);
    \item {\em Incremental}: {\tt PolyShard} allows each node to grow its local coded shard by coding over the newest verified blocks, without needing to access the previous ones. This helps to maintain a constant coding overhead as the chain grows.  
\end{itemize}

\begin{figure}[t]
  \centering
  \includegraphics[width=0.5\textwidth]{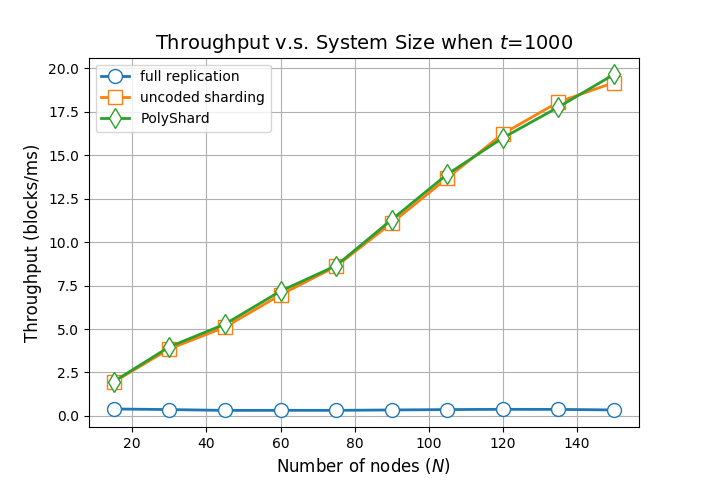}
  \caption{Measured throughput of verification schemes; here  number of epochs $t=$1000.}
  \label{fig:thpt_N_intro}
\end{figure}

As a proof of concept, we  simulate a payment blockchain system, which keeps records of balance transfers between accounts, and verifies new blocks by checking the senders' fund sufficiency. 
We run experiments on this system for various combinations of network size and chain length, and measure/evaluate the throughput, storage, and security achieved by the full replication, sharding, and the {\tt PolyShard} schemes. As we can see from the measurements plotted in Figure~\ref{fig:thpt_N_intro}, {\tt PolyShard} indeed achieves the throughput scaling with network size as the uncoded sharding scheme, improving significantly over the full replication scheme. These experiments provide an empirical  verification of the theoretical guarantees of   {\tt PolyShard} in simultaneously scaling storage efficiency, security, and throughput.

To improve the number of supported shards, we also present iterative {\tt Polyshard} in Appendix \ref{sec:iterative}. The key idea is to first represent verification functions as low-depth arithmetic circuits and then apply {\tt Polyshard} iteratively to each layer of circuits.  

In summary, the main contributions of this paper are as follows: 
\begin{itemize}[leftmargin=*]
\item Formalizing and resolving a version of the blockchain trilemma by proposing a radically different design methodology called {\tt PolyShard} that for the first time leverages coding in both storage and computation of blockchain systems.
\item Demonstrating that {\tt PolyShard} simultaneously achieves linear scaling in throughput, storage efficiency, and security; hence meeting the information-theoretic limits. 
\item Numerical evaluation of {\tt PolyShard} in a payment blockchain system and demonstrating its scalaibility in terms of throughput, storage efficiency, and security.
\end{itemize}

\noindent{\bf Other related works.} 
Prominent sharding proposals  in the literature are  \cite{luu2016secure,kokoris2017omniledger,gencer2017short,gao2017proof,zamanirapidchain,bano2017road,ren2018scale,yoo2018blockchain,cai2018decentralized,chauhan2018blockchain,manshaei2018game,EthereumSharding,gencer2016service,al2017chainspace,forestier2018blockclique}.
As an example, {\tt ELASTICO} \cite{luu2016secure} partitions the   incoming transactions into shards, and each shard is verified by a  disjoint committee of nodes in parallel.  {\tt OmniLedger}~\cite{kokoris2017omniledger} improved upon {\tt ELASTICO} in multiple avenues, including new methods to assign nodes into shards with a higher security guarantee, an atomic protocol for cross-shard transactions, and further optimization on the communication and storage designs.
\section{Problem Formulation: Block Verification}
A blockchain system manages a decentralized ledger of its clients' transactions, over a large number of untrusted network nodes. The clients submit their transactions to the network nodes, who group the transactions into blocks that to be included in the system. The accepted blocks are organized into a chain where each block contains a hash pointer to its predecessor. The chain structure provides high security guarantee since for an adversary to tamper the contents of any block, it has to re-grow the entire chain afterwards, which is extremely expensive in computation power. Here we consider a \emph{sharded} blockchain system whose grand ledger is partitioned into $K$ independent shards, each of which maintains a disjoint sub-chain that records the transactions between the accounts associated with the same shard. \footnote{We focus on transactions that are verifiable intra-shard for the sake of clarity. Extensions such as cross-shard transactions and their verification is an added complexity yet complementary to the contributions of this paper. For instance, the atomic payment and locking mechanisms of \cite{kokoris2017omniledger} can be naturally incorporated with the ideas in this paper.}
At a high level, at each time epoch, every shard proposes one block of transactions, and verifies it over the current state of its sub-chain. Once the block passes the verification, it will be appended to the corresponding sub-chain. We now define the system in more details.
\vspace{-3mm}
\subsection{Computation model}
Each shard $k$ ($k\in[1, K]$) maintains its own sub-chain. We denote the state of the $k$-th sub-chain before epoch $t$ by $Y^{t-1}_{k}=(Y_k(1),\ldots, Y_k(t-1))$, where $Y_k(t) \in \mathbb{U}$ denotes the block accepted to shard $k$ at epoch $t$, and $\mathbb{U}$ is a vector space over a field $\mathbb{F}$. When a new block $X_k(t) \in \mathbb{U}$ is proposed for shard $k$ in epoch $t$ (using some consensus algorithm like proof-of-work (PoW)), the blockchain system needs to verify its legitimacy (e.g., sender accounts have sufficient fund, no double-spending) over the current state $Y^{t-1}_{k}$. 
We abstract out the mechanism which generates the proposals and focus on verifying the proposed blocks.

We denote the verification function by $f^t: \mathbb{U}^t \rightarrow \mathbb{V}$, over $X_k(t)$ and the sub-chain $Y_k^{t-1}$, for some vector space $\mathbb{V}$ over $\mathbb{F}$. For instance, for a cryptocurrency blockchain that keeps records of balance transfers between accounts (e.g., Bitcoin), typical verification functions include 
\begin{itemize}[leftmargin=*]
    \item \emph{Balance checking}: to check each transaction has more input values than those spent in its outputs; and also check that the transactions in a block contain sufficient funds to pay the transaction fees/mining rewards.
    \item  \emph{Signature checking}: to verify that a payment transaction spending some funds in an account is indeed submitted by the owner of that account. This often involves computing cryptographic hashes using the account's public key, and verifying the results with the digital signature.
\end{itemize}
Having obtained $h_{k}^t = f^t(X_k(t), Y_k^{t-1})$, shard~$k$ computes an indicator variable $e_k^t\triangleq\mathbbm{1}(h_k^t \in \mathcal{W})$,
where $\mathcal{W} \subseteq \mathbb{V}$ denotes the set of function outputs that affirm $X_k(t)$.
Finally, the verified block $Y_k(t)$ is computed as $Y_k(t) = e_k^t X_k(t)$, and added to the sub-chain of shard $k$. We note that if the block is invalid, an all-zero block will be added at epoch $t$.

\vspace{-3mm}
\subsection{Networking model}
The above blockchain system is implemented distributedly over $N$ untrusted nodes. We consider a homogeneous and synchronous network, i.e., all nodes have similar process power, and the delay of communication between any pair of nodes is bounded by some known constant. A subset $\mathcal{M} \subset \{1,\ldots,N\}$ of the nodes may be corrupted, and are subject to Byzantine faults, i.e., they may compute and communicate arbitrary erroneous results during block verification. We aim to design secure verification schemes against the following strong adversary model:
\begin{itemize}[leftmargin=*]
    \item The adversaries can corrupt a fixed fraction of the network node, i.e., the number of malicious nodes grows linearly with $N$.\footnote{While here the adversary model is defined in a permissioned setting, the model and the proposed solution can directly extend to a permissionless setting where e.g., in a PoW system, the adversaries can control a fixed fraction of the entire hashing power.}
    \item If a conventional sharding solution were employed, the adversaries know the allocation of nodes to the shards, and are able to adaptively select the subset $M$ of nodes to attack.
\end{itemize}
We note that under this adversary model, the random shard rotation approach~\cite{luu2016secure,kokoris2017omniledger} is no longer secure since the adversaries can focus their power to attack a single shard after knowing which nodes are assigned into this shard. 
Next, we present the networking protocol that will be followed by the honest nodes. Note that adversarial nodes are not required to follow this protocol.

\noindent {\bf Storage.} At epoch $t$, each node $i$, $i=1,\ldots,N$, locally stores some data, denoted by $Z_i^{t-1} = (Z_i(1),\ldots,Z_i(t-1))$, where $Z_i(j) \in \mathbb{W}$ for some vector space $\mathbb{W}$ over $\mathbb{F}$. The locally stored data $Z_i^{t-1}$ is generated from all shards of the blockchain using some function $\phi_i^{t-1}$, i.e., $Z_i^{t-1} = \phi_i^{t-1}(Y_1^{t-1},\ldots,Y_K^{t-1})$.

\noindent {\bf Verification.} Given the $K$ proposed blocks $\{X_k(t)\}_{k=1}^K$, one for each shard, the goal of block verification is to compute $\{f^t(X_k(t),Y_k^{t-1}\}_{k=1}^K$ distributedly over untrusted nodes. We implement this verification process in two steps. In the first step, each node $i$ computes an intermediate result $g_i^t$ using some function $\rho_i^t$ on the proposed blocks and its local storage, such that $g_i^t = \rho_i^t(X_1(t),\ldots,X_K(t),Z_i^{t-1})$,
and then broadcasts the result $g_i^t$ to all other nodes. 

The nodes exploit the received computation results to reduce the final verification results in the second step. Specifically, each node $i$ decodes the verification results for all $K$ shards $\hat{h}^t_{1i},\ldots,\hat{h}^t_{Ki}$, 
using some function $\psi_i^t$, i.e., $    (\hat{h}^t_{1i},\ldots,\hat{h}^t_{Ki})= \psi_i^t(g_1^t,\ldots,g_N^t)$. 
Using these decoded results, node $i$ computes the indicator variables $\hat{e}_{ki} = \mathbbm{1}(\hat{h}_{ki} \in \mathcal{W})$, and then the verified blocks $\hat{Y}_{ki}(t) = \hat{e}^t_{ki}X_k(t)$, for all $k=1,\ldots,K$. Finally, each node~$i$ utilizes the verified blocks to update its local storage using some function $\chi_i^t$, i.e., $ Z_i^t =  \chi_i^t(\hat{Y}_{1i}(t),\ldots,\hat{Y}_{Ki}(t), Z_i^{t-1}) = \phi_i^t(\hat{Y}_{1i}^t,\ldots, \hat{Y}_{Ki}^t)$. Here $\hat{Y}_{ki}^t$ is the sequence of blocks in shard $k$ verified at node~$i$ up to time $t$. To update the local storage $Z_i^t$, while node~$i$ can always apply $\phi_i^t$ on the uncoded shards $\hat{Y}_{1i}^t,\ldots, \hat{Y}_{Ki}^t$, it is highly desirable for $\chi_i^t$ to be {\em incremental}: we only need to create a coded block from $\hat{Y}_{1i}(t),\ldots,\hat{Y}_{Ki}(t)$, and append it to $Z_i^{t-1}$. This helps to significantly reduce the compational and storage complexities. 

\subsection{Performance metrics}
We denote a block verification scheme by $S$, defined as a sequence of collections of the functions, i.e., $S = (\{\phi_i^t,\rho_i^t, \psi_i^t,\chi_i^t\}_{i=1}^N)_{i=1}^{\infty}$. We are interested in the following three performance metrics of $S$.

\noindent {\bf Storage efficiency}. 
Denoted by $\gamma_S$, it is defined as the ratio between the size of the entire block chain and the size of the data stored at each node, i.e.,
\begin{align}
    \gamma_S \triangleq \frac{K \log |\mathbb{U}|}{\log |\mathbb{W}|}.
\end{align}

The above definition also applies to a  probabilistic formulation where the blockchain elements $Y_k(j)$s and the storage elements $Z_i(j)$s are modelled as i.i.d. random variables with uniform distribution in their respective fields, where the storage efficiency is defined using the entropy of the random variables.


\noindent {\bf Security.}  We say $S$ is $b$-secure if the honest nodes can recover all the correct verification results under the presence of up to $b$ malicious nodes. More precisely, for any subset $\mathcal{M} \subset \{1,\ldots,N\}$ of malicious nodes with $|\mathcal{M}| \leq b$, and each node $i \notin \mathcal{M}$, a $b$-secure scheme will guarantee that $(\hat{h}^t_{1i},\ldots,\hat{h}^t_{Ki}) = (h^t_1,\ldots,h^t_K)$, for all $t=1,2,\ldots$.
We define the security of $S$, denoted by $\beta_S$, as the maximum $b$ it could achieve:
\begin{align}
    \beta_S \triangleq \sup\{b: \textup{$S$ is $b$-secure}\}. 
\end{align}

\noindent {\bf Throughput.} We measure throughput of the system by taking into account the number of blocks verified per epoch and the associated computational cost. We denoted by $c(f)$ the computational complexity of a function $f$, which is the number of additions and multiplications performed in the domain of $f$ to evaluate $f$.

We define the throughput of $S$, denoted by $\lambda_S$, as the average number of blocks that are correctly verified per unit discrete round, which includes all the computations performed at all $N$ nodes to verify the incoming $K$ blocks.
That is, 
\begin{align}
    \lambda_S \triangleq \liminf_{t \rightarrow \infty} \frac{K}{\sum_{i=1}^N (c(\rho_i^t) + c(\psi_i^t) + c(\chi_i^t))/(N c(f^t))}.
\end{align}



The above three metrics correspond to the three traits in the blockchain trilemma, and all current blockchain systems have to trade off one for another. The goal of this paper is to understand the information-theoretic limits on these metrics, and design verification schemes that can simultaneously achieve the limits, hence settling the trilemma.

\section{Baseline Performance}
We first present the information-theoretic {\em upper} bounds on the  three performance metrics for any blockchain. We then study the performance of two state-of-the-art blockchain schemes and comment on the gaps with respect to the upper bounds.

\noindent {\bf Information-theoretic upper bounds.} 
In terms of security, the maximum number of adversaries any verification scheme can tolerate cannot exceed half of the number of network nodes $N$. Thus, the security $\beta \leq \frac{N}{2}$.
In terms of storage, for the verification to be successful, the size of the chain should not exceed the aggregated storage resources of the $N$ nodes. Otherwise, the chain cannot be fully stored. We thus have $\gamma \leq N$.
Finally, to verify the $K$ incoming blocks, the verification function $f^t$ must be executed at least $K$ times in total. Hence, the system throughput $\lambda \leq \frac{K}{K/N} = N$.
Therefore, the information-theoretic upper bounds of security, storage efficiency, and throughput all scale linearly with the network size $N$.

\noindent{\bf Full replication. }
In terms of storage efficiency, since each node stores all the $K$ shards of the entire blockchain, full replication scheme yields $\gamma_{\text{full}}  = 1$.
Since every node verifies all the $K$ blocks, the throughput of the full replication scheme is $\lambda_{\text{full}} = \frac{K}{NKc(f^t)/(Nc(f^t))} = 1$. Thus the full replication scheme does not scale with the network size, as both the storage and the throughput remain constant as $N$ increases. The advantage is that the simple majority-rule will allow the correct verification and update of every block as long as there are less than $N/2$ malicious nodes. Thus, $\beta_{\text{full}}=N/2$.



\noindent{\bf Uncoded sharding scheme.}
In conventional {\em sharding}, the blockchain consists of $K$ disjoint sub-chains known as shards. The $N$ nodes are partitioned into $K$ groups of equal size $q = N/K$, and each group of nodes manage a single shard; this is  a full replication system with $K'=1$ shard and $N'=q$ nodes. 
Since each node stores and verifies a single shard, the storage efficiency and throughput become
$\gamma_{\textup{sharding}} = K$ and $\lambda_{\textup{sharding}} = \frac{K}{Nc^t/(Nc^t)}=K$, respectively. For these two metrics to scale linearly with $N$, it must be true that $K=\Theta(N)$. Consequently, the group size $q$ becomes a constant. Hence, compromising as few as $q/2$ nodes will corrupt one shard and the chain. Thus, this scheme only has a constant security of $\beta_{\text{sharding}}=q/2 =O(1)$. Although system solutions such as shard rotations can help achieve linearly scaling security guarantees, they are only secure when the adversary is non-adaptive (or very slowly adaptive). When the adversary is dynamic, it can corrupt all nodes belonging to a particular shard instantaneously after the shard assignment has been made. Under this model, the security reduces to a constant.

In summary, neither full replication nor the above sharding scheme exempts from the blockchain trilemma, and has to make tradeoff in scaling towards the information-theoretic limits. Motivated by the recent advances on leveraging coding theory to optimize the performance of distributed computation (see, i.e.,~\cite{li2016fundamental,LMA_all,lee2018speeding,dutta2016short,yu2017polynomial,TLDK-ICML,yu2018lagrangeNIPS}), we propose {\tt PolyShard} (polynomially coded sharding) to achieve all of the three upper bounds simultaneously. Using {\tt PolyShard}, each node stores coded data of the blockchain, and computes verification functions directly on the coded data. 


\section{PolyShard for balance checking}\label{sec:balance}

We consider a simple cryptocurrency blockchain system that records  
balance transfers between accounts. We assume that there are $M$ accounts (or addresses) associated with each shard, for some constant $M$ that does not scale with $t$. For the purpose of balance checking, we can compactly represent the block of transactions submitted to shard $k$ at epoch $t$, $X_k(t)$, as a pair of real vectors $X_k(t) = (X_k^{\textup{send}}(t), X_k^{\textup{receive}}(t)) \in \mathbb{R}^M \times \mathbb{R}^M$. 
Given a transaction in the block that reads ``Account $p$ sends $s$ units to account $q$'', we will have $X_k^{\textup{send}}(t)[p] = -s$, and $X_k^{\textup{receive}}(t)[q] = s$. Accounts that do no send/receive funds will have their entries in the send/receive vectors set to zeros. To verify $X_k(t)$, we need to check that all the sender accounts in $X_k(t)$ have accumulated enough unspent funds from previous transactions. This naturally leads to the following verification function.
\begin{align}
     f^t(X_k(t), Y_k^{t-1}) = X_k^{\textup{send}}(t) + \sum_{i=1}^{t-1} (Y_k^{\textup{send}}(i) +Y_k^{\textup{receive}}(i)). \label{eq:multiTrans_verify}
\end{align}
This function is linear in its input, and has computational complexity $c(f^t) = O(t)$.
We claim the block $X_k(t)$ valid (i.e., $e_k^t=1$) if no entry in the function's output vector is negative, and invalid (i.e., $e_k^t=0$) otherwise. After computation, we have the verified block $Y_k(t) =(Y_k^{\textup{send}}(t), Y_k^{\textup{receive}}(t))=e_k^tX_k(t)$.
We note that this balance checking can be alternatively implemented by having each shard simply store a dimension-$M$ vector that records the aggregated balances of all associated accounts, and the storage and the verification complexity will stay constant as time progresses. However, important transaction statistics including how many transactions occur in each block, and the input/output values in each transaction, and how these statistics evolve over time will be lost in this simplified implementation. Moreover, storing a single block in each shard without the protection of a long chain makes the shard more vulnerable to malicious tampering, compromising the security guarantee. Therefore, we stick to the chain structure where all past transactions are kept in the ledger.

We consider operating this sharded payment blockchain 
over a network of $N$ nodes, with a maximum $\mu$ fraction of which are corrupted by quickly-adaptive adversaries. For this system, we propose a coded transaction verification scheme, named {\tt PolyShard}, which simultaneously scales the storage efficiency, security, and throughput with the network size. 

\subsection{Coded sub-chain}
In {\tt PolyShard}, at epoch~$t$, each node $i$ stores a coded sub-chain $\tilde{Y}_i^t = (\tilde{Y}_i(1),\ldots,\tilde{Y}_i(t))$ where each component $\tilde{Y}_i(m)$ is a coded block generated by the verified blocks $Y_1(m),\ldots,Y_K(m)$ from all $K$ shards in epoch $m$. The coding is through evaluation of the well-known Lagrange interpolation polynomial. Specifically,  
we pick $K$ arbitrarily distinct real numbers $\omega_1,\ldots,\omega_K \in \mathbb{R}$, each corresponding to a shard $k$. Then for each $m=1,\ldots, t$, we create a Lagrange polynomial in variable $z$ as follows.
\begin{align}\label{eq:encoding-chain}
    u_m(z) = \sum_{k=1}^K Y_k(m) \prod_{j \neq k} \frac{z - \omega_j}{\omega_k - \omega_j}.
\end{align}
We note that this polynomial is designed such that $u_m(\omega_k) = Y_k(m)$ for all $k=1,\ldots,K$.

\begin{figure}[t]
  \centering
  \includegraphics[width=0.48\textwidth]{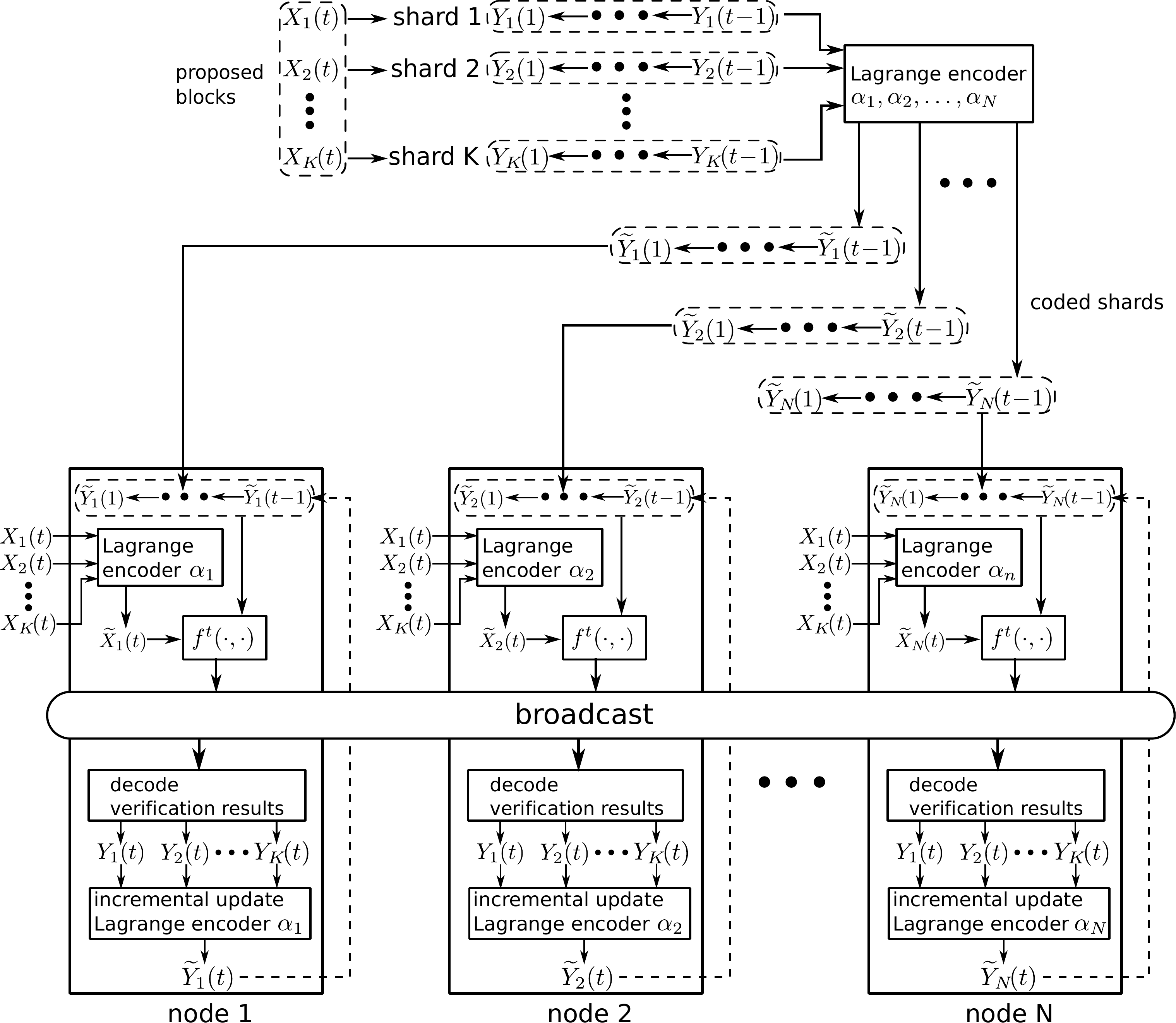}
  \caption{Illustration of {\tt PolyShard} scheme.}
  \label{fig:PolyShard}
\end{figure}

Next, as shown in Figure~\ref{fig:PolyShard}, we pick $N$ arbitrarily distinct numbers $\alpha_1,\ldots,\alpha_N \in \mathbb{R}$, one for each node. For each $i=1,\ldots,N$, we create a coded block $\tilde{Y}_i(m)$ that is stored at node~$i$, by evaluating the above $u_m(z)$ at the point $\alpha_i$. That is,
\begin{align}
\tilde{Y}_i(m) =  u_m(\alpha_i) = \sum_{k=1}^K Y_k(m) \prod_{j \neq k} \frac{\alpha_i - \omega_j}{\omega_k - \omega_j} = \sum_{k=1}^K \ell_{ik}Y_k(m).\label{eq:hisEncode}
\end{align}
We note that $\tilde{Y}_i(m)$ is a linear combination of the uncoded blocks $Y_1(m),\ldots,Y_K(m)$, and the coefficients $\ell_{ik} = \prod_{j \neq k} \frac{\alpha_i - \omega_j}{\omega_k - \omega_j}$ do not depend on the time index $m$. Therefore, one can think of each node $i$ as having a fixed vector $\ell_{ik}$ by which it mixes the different shards to create $\tilde{Y}_i^t$, and stores it locally. The size of each coded sub-chain is $K$ times smaller than the size of the entire blockchain, and the storage efficiency of {\tt PolyShard} is  $\gamma_{\textup{{\tt PolyShard}}} = K$.

\begin{remark}
The above data encoding is oblivious of the verification function, i.e., the coefficients $\ell_{ik}$ are independent of $f^t$. Therefore, the block encoding of {\tt PolyShard} can be carried out independent of the verification, and the same coded sub-chain can be simultaneously used for all other types of verification items, which could include verifying digital signatures and checking smart contracts.
\end{remark}

\subsection{Coded verification}
At epoch $t$, each shard $k$ proposes and broadcasts a new block $X_k(t)$ to be added to the sub-chain after balance checking.
{\tt PolyShard} scheme verifies these blocks in three steps.

\noindent {\bf Step 1: block encoding.} Upon receiving the $K$ proposed blocks, each node~$i$ computes a coded block $\tilde{X}_i(t)$ as a linear combination using the same set of coefficients in (\ref{eq:hisEncode}). That is, $\tilde{X}_i(t) = (\tilde{X}_i^{\textup{send}}(t), \tilde{X}_i^{\textup{receive}}(t))= \sum_{k=1}^K \ell_{ik} X_k(t)$.
We note that this encoding operation can also be viewed as evaluating the polynomial $v_t(z) = \sum_{k=1}^K X_k(t) \prod_{j \neq k} \frac{z - \omega_j}{\omega_k - \omega_j}$ at the point $\alpha_i$. This step incurs $O(NK)$ operations across the network, since each of the $N$ nodes computes a linear combination of $K$ blocks.

\noindent {\bf Step 2: coded computation.}
Each node $i$ applies the verification function 
$f^t$ in (\ref{eq:multiTrans_verify}) directly to the coded block $\tilde{X}_i(t)$, and its locally stored coded sub-chain $\tilde{Y}_i^{t-1}$ to compute an intermediate vector
\begin{align}
g_i^t =  \tilde{X}_i^{\textup{send}}(t) + \sum_{m=1}^{t-1} (\tilde{Y}_i^{\textup{send}}(m) + \tilde{Y}_i^{\textup{receive}}(m)). \label{eq:local_evaluation}
\end{align}
Each node $i$ carries out $\Theta(t)$ operations to compute $g_i^t$, and broadcasts it to all other nodes.


\noindent {\bf Step 3: decoding.}
It is easy to see that $f^t(v_t(z),u_1(z),\ldots,u_{t-1}(z))$ is a univariate polynomial of degree $K-1$, and $g_i^t$ can be viewed as evaluating this polynomial at $\alpha_i$. Given the evaluations at $N$ distinct points $g_1^t,\ldots,g_N^t$, each node can recover the coefficients of $f^t(v_t(z),u_1(z),\ldots,u_{t-1}(z))$ following the process of decoding a Reed-Solomon code with dimension $K$ and length $N$ (see, e.g.,~\cite{roth2006introduction}). In order for this decoding to be robust to $\mu N$ erroneous results (i.e., achieving security $\beta_{\textup{{\tt PolyShard}}} = \mu N$), we must have $2\mu N \leq N-K$. In other words, a node can successfully decode $f^t(v_t(z),u_1(z),\ldots,u_{t-1}(z))$ only if the number of shards $K$ is upper bounded by $K \leq (1-2\mu)N$. 
Based on this constraint, we set the number of shards of the {\tt PolyShard} scheme, $K_{\textup{{\tt PolyShard}}} = \lfloor (1-2\mu)N \rfloor$, which scales linearly with network size $N$.

The complexity of decoding a length-$N$ Reed-Solomon code at each node is $O(N \log^2 N \log \log N)$, and the total complexity of the decoding step is $O(N^2 \log^2 N \log \log N)$.

Having decoded $f^t(v_t(z),u_1(z),\ldots,u_{t-1}(z))$, each node evaluates it at $\omega_1,\ldots,\omega_K$ to recover $\{f^t(X_k(t),Y_k^{t-1})\}_{k=1}^K$, to obtain the verification results $\{e_k^t\}_{k=1}^K$ and the verified blocks $\{Y_k(t) = e_k^t X_k(t)\}_{k=1}^K$. Finally, each node $i$ computes $\tilde{Y}_i(t)$ following (\ref{eq:hisEncode}), and appends it to its local coded sub-chain. Updating the sub-chains has the same computational complexity with the block encoding step, which is $O(NK)$.

\begin{remark}
The sub-chain update process in {\tt PolyShard} is incremental, i.e., appending a coded block to a coded sub-chain is equivalent to appending uncoded blocks to the uncoded sub-chains, and then encoding from the updated sub-chains. This commutativity between sub-chain growth and shard encoding allows each node to update its local sub-chain incrementally by accessing only the newly verified blocks instead of the entire block history.  
\end{remark}

\subsection{Performance of {\tt PolyShard}}\label{sec:performance}
So far, we have shown that {\tt PolyShard} achieves a storage efficiency $\gamma_{\textup{{\tt PolyShard}}} = K_{\textup{{\tt PolyShard}}} = \Theta(N)$, and it is also robust against $\mu N = \Theta(N)$ quickly-adaptive adversaries. The total number of operations during the verification and the sub-chain update processes is
$O(NK) + N\Theta(t) + O(N^2 \log^2 N \log\log N)$, where the term $O(NK) + O(N^2 \log^2 N \log\log N)$ is the additional coding overhead compared with the uncoded sharding scheme. Since $K_{\textup{{\tt PolyShard}}}\leq N$,
the coding overhead 
reduces to $O(N^2 \log^2 N \log\log N)$. The throughput of {\tt PolyShard} for balance checking is computed as
\begin{align}
    \lambda_{\textup{{\tt PolyShard}}} &= \liminf_{t \rightarrow \infty} \frac{K_{\textup{{\tt PolyShard}}}N\Theta(t)}{N\Theta(t) + O(N^2 \log^2 N \log\log N)}\\
    &=\liminf_{t \rightarrow \infty} \frac{K_{\textup{{\tt PolyShard}}}}{1 + \frac{O(N\log^2 N \log \log N)}{\Theta(t)}}= \Theta(N). 
\end{align}

We can see that since the complexities of the encoding and decoding operations of {\tt PolyShard} do not scale with $t$, the coding overhead becomes irrelevant as the chain grows. The {\tt PolyShard} scheme simultaneously achieves information-theoretically optimal scaling on security, storage efficiency, and throughput.

We note that when the verification function is linear with the block data, codes designed for distributed storage (see, e.g.,~\cite{dimakis2011survey,rashmi2011optimal}) can be used to achieve similar scaling as {\tt PolyShard}. However, {\tt PolyShard} is designed for a much more general class of verification functions including arbitrary multivariate polynomials, which cannot be handled by state-of-the-art storage codes.

\section{PolyShard for general verification functions}\label{sec:general}

In this section, we describe the {\tt PolyShard} scheme for a more general class of verification functions $f^t$ that can be represented as a multivariate polynomial with maximum degree of $d$. Cryptographic hash functions, which are extensively used in computing account addresses, validating transaction Merkle trees, and verifying digital signatures, are often evaluated as polynomials of its input data. For example, the Z\'emor-Tillich hash function is computed by multiplying $2\times 2$ matrices in a finite field of characteristic $2$, whose degree is proportional to the number of input bits~\cite{tillich1994hashing}. On the other hand, for hash functions that are based on bit mixing, and often lack algebraic structures (e.g., SHA-2, and Keccak~\cite{Keccak}), we can represent the corresponding Boolean verification functions (indicating whether the block is valid or not) as polynomials using the following result~\cite{zou2011representing}: any Boolean function $\{0,1\}^n \rightarrow \{0,1\}$ can be represented by a polynomial of degree $\leq n$ with at most $2^{n-1}$ terms. The explicit construction of this polynomial is described in Appendix~\ref{sec:field}.  

As one of the main advantages of {\tt PolyShard}, the system design is almost independent of the verification functions, and the {\tt PolyShard} scheme for high-degree ($d \geq 1$) polynomials is almost identical to that of balance checking. In particular, at epoch~$t$, the latest block stored at node~$i$, i.e., $\tilde{Y}_i(t-1)$, is generated as in~(\ref{eq:hisEncode}). In contrast to the case of balance checking that operates over real numbers, when the underlying field $\mathbb{F}$ is finite, we will need the field size $|\mathbb{F}|$ to be at least $N$ for this block encoding to be viable. For small field (e.g., binary field), we can overcome this issue by using field extension and applying {\tt PolyShard} on the extended field (see details in Appendix~\ref{sec:field}).

The block verification process is similar to that of the balance checking example. Having created the coded block $\tilde{X}_i(t)$, node~$i$ directly applies the verification function to compute $g_i^t = f^t(\tilde{X}_i(t),\tilde{Y}_i^{t-1})$, and broadcasts it to all other nodes. Since $f^t$ is a polynomial of degree $d$, $f^t(v_t(z),u_1(z),\ldots,u_{t-1}(z))$ becomes a polynomial of degree $(K-1)d$. Now to decode $f^t(v_t(z),u_1(z),\ldots,u_{t-1}(z))$ from $g_1^t,\ldots,g_N^t$, each node needs to decode a Reed-Solomon code with dimension $(K-1)d+1$ and length $N$.
Successful decoding would require the number of errors $\mu N \leq (N-(K-1)d-1)/2$. That is, the maximum number of shards $K$ that can be securely supported is $K_{\textup{{\tt PolyShard}}} = \lfloor \frac{(1-2\mu)N-1}{d}+1 \rfloor$.

While it is fairly clear that {\tt PolyShard} achieves storage efficiency $\gamma_{\textup{{\tt PolyShard}}} =K_{\textup{{\tt PolyShard}}}= \Theta(N)$ and security $\beta_{\textup{{\tt PolyShard}}} = \mu N = \Theta(N)$ for a dergee-$d$ verification function, the throughput of {\tt PolyShard} is
\begin{align}
      \lambda_{\textup{{\tt PolyShard}}}= \liminf_{t \rightarrow \infty} \frac{K_{\textup{{\tt PolyShard}}}}{1 + \frac{O(N \log^2 N \log\log N)}{c(f^t)}}.  \label{eq:T_PS}
\end{align}
When $c(f^t)$ grows with $t$, (e.g., $c(f^t)=\Theta(t)$ for the above balance checking function that scans the entire sub-chain to find sufficient funds), the throughput of ${\tt PolyShard}$ becomes $ \lambda_{\textup{{\tt PolyShard}}} = K_{\textup{{\tt PolyShard}}}=\Theta(N)$. We summarize the scaling results of the {\tt PolyShard} scheme in the following theorem.

\begin{theorem}\label{thm:PolyShard}
Consider a sharded blockchain whose blocks in each shard are verified by computing a multivariate polynomial of degree $d$ on the blocks in that shard. When implementing this blockchain over $N$ network nodes, up to $\mu$ (for some constant $0 \leq \mu < \frac{1}{2}$) fraction of which may be corrupted by quickly-adptive adversaries, the proposed ${\tt PolyShard}$ scheme supports secure block verification from up to $K_{\textup{{\tt PolyShard}}} = \lfloor \frac{(1-2\mu)N-1}{d}+1 \rfloor = \Theta(N)$ shards, and simultaneously achieves the following performance metrics,
\begin{align*}
  \textup{Storage efficiency} \quad  \gamma_{\textup{{\tt PolyShard}}}&=   K_{\textup{{\tt PolyShard}}}=\Theta(N),\\
  \textup{Security} \quad   \beta_{\textup{{\tt PolyShard}}} &= \mu N = \Theta(N),\\
   \textup{Throughput} \quad   \lambda_{\textup{{\tt PolyShard}}} &= K_{\textup{{\tt PolyShard}}}=\Theta(N),
\end{align*}
when computational complexity of the verification function grows with the length of the sub-chain in each shard. Therefore, {\tt PolyShard} simultaneously achieves the information-theoretically optimal storage efficiency, security, and throughput to within constant multiplicative gaps.
\end{theorem}
The number of shards supported by {\tt Polyshard} decreases as degree of polynomial increases. One can remove such limitation by using {\tt Polyshard} iteratively. The main idea of iterative {\tt Polyshard} is to represent the verification function as a low-depth arithmetic circuit which can be implemented iteratively by computing low-degree polynomials, and then apply {\tt Polyshard} to the low-degree polynomial iteratively. We show that the number of supported shards can be independent of the degree of function by the following theorem. (see details in Appendix \ref{sec:iterative}).
\begin{theorem}\label{thm:IterPolyShard}
Consider a sharded blockchain whose blocks in each shard are verified by computing a multivariate polynomial of degree $d$ on the blocks in that shard. When implementing this blockchain over $N$ network nodes, up to $\mu$ (for some constant $0 \leq \mu < \frac{1}{2}$) fraction of which may be corrupted by quickly-adptive adversaries, the proposed iterative ${\tt PolyShard}$ scheme supports secure block verification from up to $K_{\textup{{\tt Iterative}}} = \lfloor \frac{(1-2\mu)N+1}{2} \rfloor$.
\end{theorem}
\begin{remark}
The shard encoding and computation decoding schemes of {\tt PolyShard} are developed based on the coded computation techniques proposed in~\cite{yu2018lagrangeNIPS}, which are used for distributed computing multivariate polynomials subject to arbitrary computation errors. Speficically, it is proposed in~\cite{yu2018lagrangeNIPS} to create coded data batches using Lagrange polynomial interpolation, and perform computations directly on coded data. However, in contrast to the scenario of one-shot computation on static data in~\cite{yu2018lagrangeNIPS}, the locally stored data at each node is growing in a blockchain system, as more verified blocks are appended to the chain. The requirement of dynamically updating the local coded sub-chain that is compatible with the upcoming coded verification poses new challenges on the design of the {\tt PolyShard} scheme. Utilizing the data structure of the blockchain, and the algebraic properties of the encoding strategy, we propose a simple \emph{incremental} sub-chain update policy for {\tt PolyShard} that requires accessing the minimum amount of data. 
\end{remark}

\begin{remark}
The additional coding overhead, including the operations required to encode the incoming blocks, decode verification results, and update the coded shards, does not scale with the length of the sub-chain $t$. 
When the complexity of computing $f^t$, i.e., $c(f^t)$, grows with $t$, the coding overhead becomes negligible as the chain grows, and the throughput of {\tt PolyShard} scales linearly with the network size. However, on the other hand, when $c(f^t)$ is independent of chain length (e.g., verifying digital signature that only requires data from the current block), the coding overhead will dominate the local computation. In this case, while {\tt PolyShard} still achieves scalability on storage and security, its throughput remains constant as the network grows.
\end{remark}

\section{Simulation Results}\label{sec:simulation}
\def\fr{full replication}
\def\Fr{Full replication}
\def\ucs{uncoded sharding}
\def\Ucs{Uncoded sharding}
\def\PS{{\tt PolyShard~}}
We perform detailed simulations to assess the performance of {\tt PolyShard} for balance checking described in Section~\ref{sec:balance}. The blockchain system keeps records of all the balance transfers between accounts, and verifies new blocks by comparing them with the sum of the previously verified blocks (i.e., computing the verification function in (\ref{eq:multiTrans_verify})). More specifically, the system contains $K$ shards, each managing $M$ accounts. At each time epoch $t$, each shard $k$ proposes a block of transactions for verification. 
On a single computer, we simulate this blockchain system over $N$ network nodes, using full replication, uncoded sharding, and {\tt PolyShard} schemes respectively. During the simulation, we execute a serial program that performs the local computations of the nodes one after another, and measure each of these computation times in serial. All node instances share the same memory, so the communication delay between nodes is negligible. 

We compute the throughput of each scheme under different values of $N$ and $t$ to understand its scalability. Throughput is defined as the number of blocks verified per time unit, and is computed by dividing $K$ (the number of blocks proposed per epoch) by the average computation time (measured during simulation) of the $N$ nodes. For {\tt PolyShard}, the computation time also includes the time each node spends on encoding the blocks. However, since the encoding time is a constant, whilst the balance summation time increases with $t$ as the chain becomes longer, it is expected that the encoding time is becoming negligible. We note that the storage efficiency and security level of each scheme are decided by system parameters and, thus, do not need measurements.

\begin{figure}[t]
  \centering
  \includegraphics[width=0.49\textwidth]{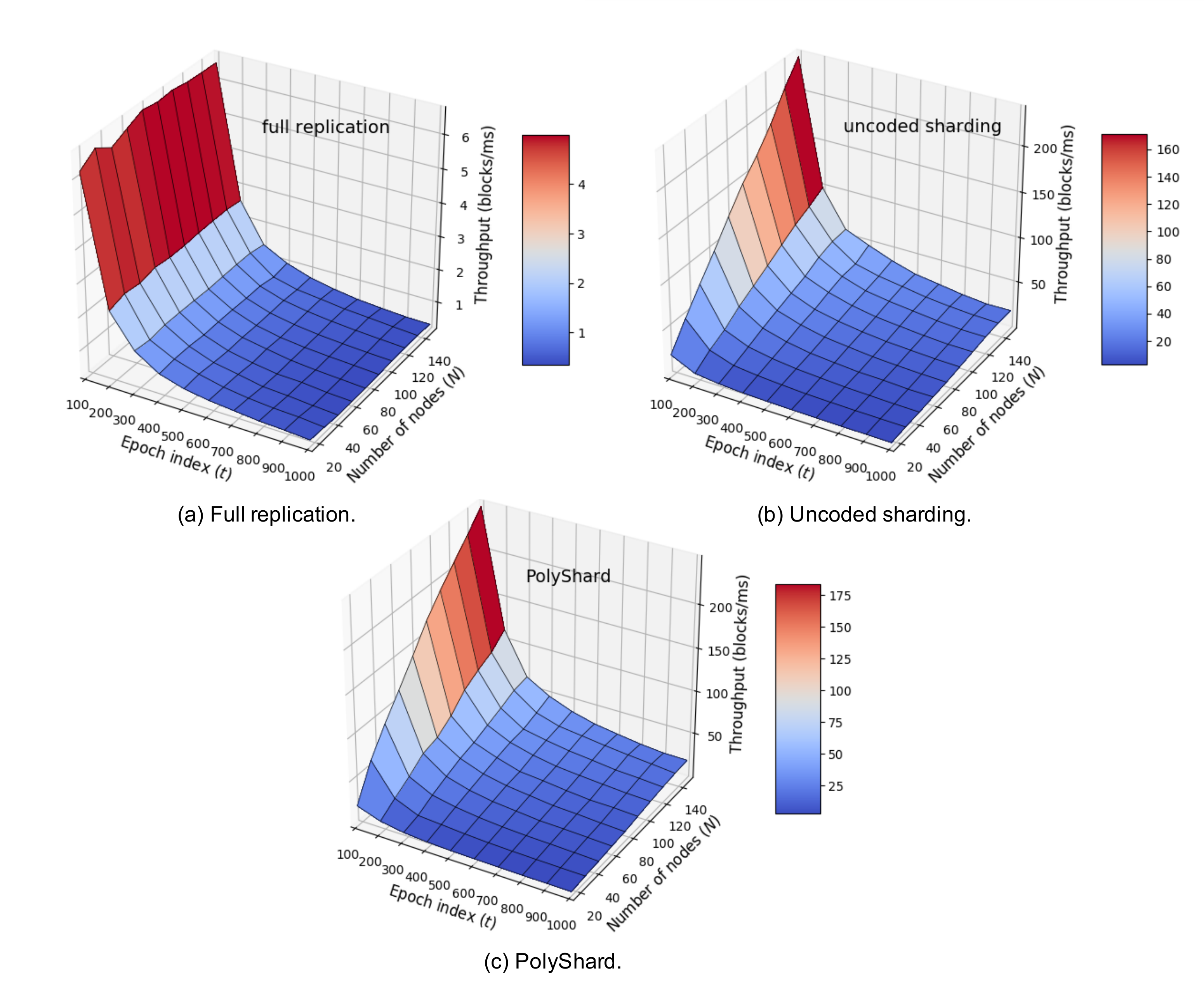}
  \caption{Throughput of the three schemes with respect to time and number of nodes.}
  \label{fig:thpt_all}
\end{figure}

\begin{figure}[t]
  \centering
  \includegraphics[width=0.48\textwidth]{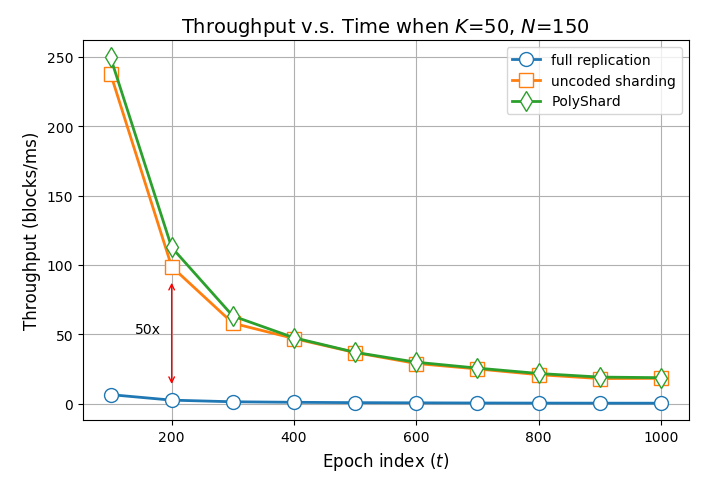}
  \caption{Throughput of the three schemes when number of nodes $N=$150.}
  \label{fig:thpt_epoch}
\end{figure}

We simulate this system for $t=1000$ epochs, using different number of shards $K\in[5, 50]$. Each shard manages $M=2000$ accounts. We fix the ratio $N/K=3$. Thus, the number of nodes is $N\in[15, 150]$. We plot the complete relationship between $N$, $t$, and throughput of the three schemes in Figure~\ref{fig:thpt_all}. For a closer look, 
We plot the relationship between the sub-chain length $t$ and throughput when $N=150$ in Figure~\ref{fig:thpt_epoch}, and the relationship between the network size $N$ and throughput when $t=1000$ in Figure~\ref{fig:thpt_N_intro} in Section~\ref{sec:intro}.

\vspace{3mm}
\noindent {\bf Results and discussions}
\vspace{1mm}
\begin{enumerate}[leftmargin=*]
    \item Throughput: As expected, {\tt PolyShard} provides the same throughput as \ucs, which is about $K$ times of the throughput of \fr. From Figure~\ref{fig:thpt_epoch}, we observe that the throughput of all three schemes drops as the time progresses. This is because that the computational complexity of verifying a block increases as more blocks are appended to each shard. In terms of scalability, Fig.~\ref{fig:thpt_N_intro} indicates that the throughput of {\tt PolyShard} and \ucs~both increases linearly with the network size $N$ (and $K$), whilst the throughput of \fr~almost stays the same. 
    
    \item Storage: It is straightforward to see that {\tt PolyShard} provides the same storage gain over \fr~as \ucs, with a factor of $K$. Thus, {\tt PolyShard} and \ucs~are scalable in storage, but \fr~is not (Table~\ref{table:compare_storage}).
    
    \item Security: As we have analyzed, \fr~can tolerate up to 50\% of malicious nodes, achieving the maximum security level $\beta_{\textup{full}}=\frac{N}{2}$. The error-correcting process of {\tt PolyShard} provides robustness to $\beta_{{\tt PolyShard}} = \frac{N-K}{2} = \frac{N-N/3}{2} = \frac{N}{3}$ malicious nodes. In contrast, under \ucs, each shard is only managed by $3$ nodes. Thus, its security level is only $1$ regardless of $N$, which is not scalable (Table~\ref{table:compare_security}).
\end{enumerate}
\begin{table}[h]
\centering
\subfloat[Subtable 1 list of tables text][Storage efficiency.]{
\begin{tabular}{| c | c | c | c | c | c | c|}
    \hline
    \rule{0pt}{10pt} $N$ & 15 & 30 & 60 & 90 & 120 & 150  \\ \hline
    \rule{0pt}{10pt} $\gamma_{\textup{full}}$ & 1& 1  & 1  &  1 & 1 & 1\\ \hline
    \rule{0pt}{10pt} $\gamma_{\textup{sharding}}$ & 5 & 10 & 20  & 30 & 40 & 50\\ \hline
    \rule{0pt}{10pt} $\gamma_{\textup{{\tt PolyShard}}}$ & 5 & 10 & 20  & 30 & 40 & 50\\ \hline
  \end{tabular}\label{table:compare_storage}}
\qquad 
\subfloat[Subtable 2 list of tables text][Security.]{
\begin{tabular}{| c | c | c | c | c | c | c |}
    \hline
    \rule{0pt}{10pt} $N$ & 15 & 30 & 60 & 90 & 120 & 150  \\ \hline
    \rule{0pt}{10pt} $\beta_{\textup{full}}$ & 7 & 15 & 30 & 45 & 60 & 75\\ \hline
    \rule{0pt}{10pt} $\beta_{\textup{sharding}}$ & 1 & 1 & 1  & 1 & 1 & 1\\ \hline
    \rule{0pt}{10pt} $\beta_{\textup{{\tt PolyShard}}}$ & 5 & 10 & 20  & 30 & 40 & 50\\ \hline
  \end{tabular}\label{table:compare_security}}
\caption{Storage efficiency and security of the three schemes under different network size $N$.}
\label{table:compare_scaling}
\end{table}

In summary, {\tt PolyShard} outperforms both \fr~and \ucs~because it is the only scheme that can simultaneously 1) alleviate the storage load at each node; and 2) boost the verification throughput by scaling out the system, and 3) without sacrificing the safety requirement even when the number of adversaries also grows with network size.

\section{Discussion}
In this section, we discuss how {\tt PolyShard}  fits into the overall architecture of a contemporary blockchain system.  


\noindent {\bf Integration into blockchain systems.}
We note that {\tt Polyshard} has so far been described in a simple setting where each shard produces one block in lock-step. We highlight one instantiation of how {\tt Polyshard} could fit into the architecture of an existing blockchain system, which combines a standard sharding method for proposal followed by {\tt Polyshard} for finalization. The $K$ shards are obtained by assigning users to shards via a user-assignment algorithm. The $N$ nodes are partitioned into $K$ shards using a standard sharding system (see \cite{kokoris2017omniledger}). Inside of the shard, the nodes run a standard blockchain along with a finalization algorithm to get a {\em locally finalized} version of the block. 

Each node is also assigned a coded shard via a coded-shard-assignment algorithm, which assigns a random field element $\alpha_i \in \mathbb{F}$ to a node so that the node can compute which linear combination it will use for coding. We point out here that in a permissionless setting, it is easy to handle churn (users joining and leaving) by this method if the size of the finite field $\mathbb{F}$ is much larger than $N$ - since at this point, the probability of collision (two users getting assigned the same field element) becomes negligible. Thus each node plays a role in both an uncoded shard as well as a coded shard, thus its storage requirement will be doubled; however, our system still has storage efficiency scaling with $N$. The   {\tt Polyshard} algorithm now gets the locally finalized blocks from the different shards at regular intervals and it acts as a global finalization step performing coded validation at the level of the locally finalized blocks. We point out that users requiring high trust should wait for this global finalization stamp before confirming a payment, whereas users requiring short latency can immediately utilize the local-finalization for confirmation. 

Beyond the aforementioned issues, there may be cross-shard transactions present in the system, which are payments or smart contracts with inputs and outputs distributed across multiple shards. In such a case, we will use a locking-based method, which locks the payment at the source shard and produces a certificate to the destination shard so that the amount can be spent; this idea has been proposed as well as implemented in {\tt Elastico} \cite{luu2016secure} and {\tt Omniledger} \cite{kokoris2017omniledger}.

\noindent {\bf Relationship to verifiable computing.} An alternative paradigm for accelerating computing in blockchain is verifiable computing \cite{gennaro2010non,bitansky2012extractable,parno2016pinocchio,ben2014succinct,ben2018scalable}, where a single node executes a set of computations (for example, payment validation) and integrity of these computations are then cryptographically certified. A major difference between our framework and verifiable computing is that our scheme is {\em information-theoretically secure} against a computationally unbounded adversary as against the computational security offered by verifiable computing schemes. However, verifiable computing schemes can provide zero-knowledge proofs, whereas our scheme does not offer zero-knowledge capability. Finally, verifiable computing is relevant in an  asymmetric setting, where one computer is much more powerful than the others, unlike {\tt Polyshard} that is designed for a symmetric setup comprising of equally powerful and decentralized nodes.

\noindent {\bf Future research directions.}
{\tt Polyshard} currently works with polynomials whose degree scales sub-linearly with the number of nodes. An interesting direction of future work is to remove this limitation. In particular, computations that can be represented as {\em low-depth} arithmetic circuits can be implemented iteratively using low-degree polynomials.  Another important direction of future research is the design of validation schemes that can be represented as low-degree polynomials or low-depth arithmetic circuits.

\bibliographystyle{ieeetr}
\bibliography{polyshard.bib}

\begin{thebibliography}{10}

\bibitem{nakamoto2008bitcoin}
S.~Nakamoto, ``Bitcoin: A peer-to-peer electronic cash system,'' 2008.

\bibitem{bahga2016blockchain}
A.~Bahga and V.~K. Madisetti, ``Blockchain platform for industrial internet of
  things,'' {\em Journal of Software Engineering and Applications}, vol.~9,
  no.~10, p.~533, 2016.

\bibitem{mettler2016blockchain}
M.~Mettler, ``Blockchain technology in healthcare: The revolution starts
  here,'' in {\em IEEE 18th International Conference on e-Health Networking,
  Applications and Services (Healthcom)}, pp.~1--3, IEEE, 2016.

\bibitem{croman2016scaling}
K.~Croman, C.~Decker, I.~Eyal, A.~E. Gencer, A.~Juels, A.~Kosba, A.~Miller,
  P.~Saxena, E.~Shi, E.~G. Sirer, {\em et~al.}, ``On scaling decentralized
  blockchains,'' in {\em International Conference on Financial Cryptography and
  Data Security}, pp.~106--125, Springer, 2016.

\bibitem{trilemma}
T.~Ometoruwa, ``Solving the blockchain trilemma: Decentralization, security \&
  scalability.''
  \url{https://www.coinbureau.com/analysis/solving-blockchain-trilemma/}, 2018.
\newblock Accessed: 2018-12-21.

\bibitem{wood2014ethereum}
G.~Wood, ``Ethereum: A secure decentralised generalised transaction ledger,''
  {\em Ethereum project yellow paper}, vol.~151, pp.~1--32, 2014.

\bibitem{gervais2016security}
A.~Gervais, G.~O. Karame, K.~W{\"u}st, V.~Glykantzis, H.~Ritzdorf, and
  S.~Capkun, ``On the security and performance of proof of work blockchains,''
  in {\em Proceedings of the 2016 ACM SIGSAC Conference on Computer and
  Communications Security}, pp.~3--16, ACM, 2016.

\bibitem{luu2016secure}
L.~Luu, V.~Narayanan, C.~Zheng, K.~Baweja, S.~Gilbert, and P.~Saxena, ``A
  secure sharding protocol for open blockchains,'' in {\em Proceedings of the
  2016 ACM SIGSAC Conference on Computer and Communications Security},
  pp.~17--30, ACM, 2016.

\bibitem{kokoris2017omniledger}
E.~Kokoris-Kogias, P.~Jovanovic, L.~Gasser, N.~Gailly, and B.~Ford,
  ``Omniledger: A secure, scale-out, decentralized ledger.,'' {\em IACR
  Cryptology ePrint Archive}, vol.~2017, p.~406, 2017.

\bibitem{gencer2017short}
A.~E. Gencer, R.~van Renesse, and E.~G. Sirer, ``Short paper: Service-oriented
  sharding for blockchains,'' in {\em International Conference on Financial
  Cryptography and Data Security}, pp.~393--401, Springer, 2017.

\bibitem{LMA_all}
S.~Li, M.~A. Maddah-Ali, and A.~S. Avestimehr, ``Coded {M}ap{R}educe,'' {\em
  53rd Allerton Conference}, Sept. 2015.

\bibitem{li2016fundamental}
S.~Li, M.~A. Maddah-Ali, Q.~Yu, and A.~S. Avestimehr, ``A fundamental tradeoff
  between computation and communication in distributed computing,'' {\em IEEE
  Transactions on Information Theory}, vol.~64, Jan. 2018.

\bibitem{lee2018speeding}
K.~Lee, M.~Lam, R.~Pedarsani, D.~Papailiopoulos, and K.~Ramchandran, ``Speeding
  up distributed machine learning using codes,'' {\em IEEE Transactions on
  Information Theory}, vol.~64, no.~3, pp.~1514--1529, 2018.

\bibitem{LMA16_unify}
S.~Li, M.~A. Maddah-Ali, and A.~S. Avestimehr, ``A unified coding framework for
  distributed computing with straggling servers,'' {\em IEEE Workshop on
  Network Coding and Applications}, Sept. 2016.

\bibitem{dutta2016short}
S.~Dutta, V.~Cadambe, and P.~Grover, ``Short-dot: Computing large linear
  transforms distributedly using coded short dot products,'' in {\em NIPS},
  pp.~2100--2108, 2016.

\bibitem{yu2017polynomial}
Q.~Yu, M.~A. Maddah-Ali, and A.~S. Avestimehr, ``Polynomial codes: an optimal
  design for high-dimensional coded matrix multiplication,'' in {\em NIPS},
  pp.~4406--4416, 2017.

\bibitem{karakus2017straggler}
C.~Karakus, Y.~Sun, S.~Diggavi, and W.~Yin, ``Straggler mitigation in
  distributed optimization through data encoding,'' in {\em NIPS},
  pp.~5440--5448, 2017.

\bibitem{TLDK-ICML}
R.~Tandon, Q.~Lei, A.~G. Dimakis, and N.~Karampatziakis, ``Gradient coding:
  Avoiding stragglers in distributed learning,'' in {\em Proceedings of the
  34th International Conference on Machine Learning}, pp.~3368--3376, Aug.
  2017.

\bibitem{yu2018lagrangeNIPS}
Q.~Yu, S.~Li, N.~Raviv, S.~M.~M. Kalan, M.~Soltanolkotabi, and A.~S.
  Avestimehr, ``Lagrange coded computing: Optimal design for resiliency,
  security, and privacy,'' in {\em NIPS Systems for ML Workshop}, 2018.

\bibitem{gao2017proof}
Y.~Gao and H.~Nobuhara, ``A proof of stake sharding protocol for scalable
  blockchains,'' {\em Proceedings of the Asia-Pacific Advanced Network},
  vol.~44, pp.~13--16, 2017.

\bibitem{zamanirapidchain}
M.~Zamani, M.~Movahedi, and M.~Raykova, ``Rapidchain: A fast blockchain
  protocol via full sharding,'' {\em Cryptology ePrint Archive}, 2018.
\newblock \url{https://eprint.iacr.org/2018/460.pdf}.

\bibitem{bano2017road}
S.~Bano, M.~Al-Bassam, and G.~Danezis, ``The road to scalable blockchain
  designs,'' {\em USENIX; login: magazine}, 2017.

\bibitem{ren2018scale}
Z.~Ren and Z.~Erkin, ``A scale-out blockchain for value transfer with
  spontaneous sharding,'' {\em e-print arXiv:1801.02531}, 2018.

\bibitem{yoo2018blockchain}
H.~Yoo, J.~Yim, and S.~Kim, ``The blockchain for domain based static
  sharding,'' in {\em 2018 17th IEEE International Conference On Trust,
  Security And Privacy In Computing And Communications/12th IEEE International
  Conference On Big Data Science And Engineering (TrustCom/BigDataSE)},
  pp.~1689--1692, IEEE, 2018.

\bibitem{cai2018decentralized}
S.~Cai, N.~Yang, and Z.~Ming, ``A decentralized sharding service network
  framework with scalability,'' in {\em International Conference on Web
  Services}, pp.~151--165, Springer, 2018.

\bibitem{chauhan2018blockchain}
A.~Chauhan, O.~P. Malviya, M.~Verma, and T.~S. Mor, ``Blockchain and
  scalability,'' in {\em 2018 IEEE International Conference on Software
  Quality, Reliability and Security Companion (QRS-C)}, pp.~122--128, IEEE,
  2018.

\bibitem{manshaei2018game}
M.~H. Manshaei, M.~Jadliwala, A.~Maiti, and M.~Fooladgar, ``A game-theoretic
  analysis of shard-based permissionless blockchains,'' {\em e-print
  arXiv:1809.07307}, 2018.

\bibitem{EthereumSharding}
``Ethereum sharding {F}{A}{Q}s.''
  \url{https://github.com/ethereum/wiki/wiki/Sharding-FAQs}.

\bibitem{gencer2016service}
A.~E. Gencer, R.~van Renesse, and E.~G. Sirer, ``Service-oriented sharding with
  aspen,'' {\em e-print arXiv:1611.06816}, 2016.

\bibitem{al2017chainspace}
M.~Al-Bassam, A.~Sonnino, S.~Bano, D.~Hrycyszyn, and G.~Danezis, ``Chainspace:
  A sharded smart contracts platform,'' {\em e-print arXiv:1708.03778}, 2017.

\bibitem{forestier2018blockclique}
S.~Forestier, ``Blockclique: scaling blockchains through transaction sharding
  in a multithreaded block graph,'' {\em e-print arXiv:1803.09029}, 2018.

\bibitem{roth2006introduction}
R.~Roth, {\em Introduction to coding theory}.
\newblock Cambridge University Press, 2006.

\bibitem{dimakis2011survey}
A.~G. Dimakis, K.~Ramchandran, Y.~Wu, and C.~Suh, ``A survey on network codes
  for distributed storage,'' {\em Proceedings of the IEEE}, vol.~99, no.~3,
  pp.~476--489, 2011.

\bibitem{rashmi2011optimal}
K.~V. Rashmi, N.~B. Shah, and P.~V. Kumar, ``Optimal exact-regenerating codes
  for distributed storage at the msr and mbr points via a product-matrix
  construction,'' {\em IEEE Transactions on Information Theory}, vol.~57,
  no.~8, pp.~5227--5239, 2011.

\bibitem{tillich1994hashing}
J.-P. Tillich and G.~Z{\'e}mor, ``Hashing with $sl_2$,'' in {\em Annual
  International Cryptology Conference}, pp.~40--49, Springer, 1994.

\bibitem{Keccak}
G.~Bertoni, J.~Daemen, M.~Peeters, G.~Van~Assche, and R.~Van~Keer, ``Keccak
  specifications summary.''
  \url{https://keccak.team/keccak_specs_summary.html}.
\newblock Accessed: 2018-12-21.

\bibitem{zou2011representing}
Y.~M. Zou, ``Representing boolean functions using polynomials: more can offer
  less,'' in {\em International Symposium on Neural Networks}, pp.~290--296,
  Springer, 2011.

\bibitem{gennaro2010non}
R.~Gennaro, C.~Gentry, and B.~Parno, ``Non-interactive verifiable computing:
  Outsourcing computation to untrusted workers,'' in {\em Annual Cryptology
  Conference}, pp.~465--482, Springer, 2010.

\bibitem{bitansky2012extractable}
N.~Bitansky, R.~Canetti, A.~Chiesa, and E.~Tromer, ``From extractable collision
  resistance to succinct non-interactive arguments of knowledge, and back
  again,'' in {\em Proceedings of the 3rd Innovations in Theoretical Computer
  Science Conference}, pp.~326--349, ACM, 2012.

\bibitem{parno2016pinocchio}
B.~Parno, J.~Howell, C.~Gentry, and M.~Raykova, ``Pinocchio: Nearly practical
  verifiable computation,'' {\em Communications of the ACM}, vol.~59, no.~2,
  pp.~103--112, 2016.

\bibitem{ben2014succinct}
E.~Ben-Sasson, A.~Chiesa, E.~Tromer, and M.~Virza, ``Succinct non-interactive
  zero knowledge for a von neumann architecture.,'' in {\em USENIX Security
  Symposium}, pp.~781--796, 2014.

\bibitem{ben2018scalable}
E.~Ben-Sasson, I.~Bentov, Y.~Horesh, and M.~Riabzev, ``Scalable, transparent,
  and post-quantum secure computational integrity,'' {\em Cryptol. ePrint
  Arch., Tech. Rep}, vol.~46, p.~2018, 2018.

\end{thebibliography}
\appendices
\section{Field extension for general Boolean functions}\label{sec:field}
For general blockchain systems that verify incoming blocks based on the most recent $0 \leq P \leq t-1$ verified blocks, 
we can generally model each incoming block $X_k(t)$, and each verified block $Y_k(m)$ as a binary bit stream of length $T$, and the verification function $f^t: \{0,1\}^{T(P+1)} \rightarrow \{0,1\}$ as a Boolean function that indicates whether $X_k(t)$ is valid or not. 

Using the construction of~\cite[Theorem 2]{zou2011representing}, we can represent any arbitrary Boolean function $f: \{0,1\}^{n} \rightarrow \{0,1\}$ whose inputs are $n$ binary variables as a multivariate polynomial $p$ of degree $n$ as follows. For each vector $\mathbf{a} = (a_1,\ldots,a_n) \in \{0,1\}^n$, we define $h_{\mathbf{a}} = z_1z_2\cdots z_n$, where $z_i = x_i$ if $a_i=1$, and $z_i=y_i$ if $a_i=0$. Next, we partition $\{0,1\}^n$ into two disjoint subsets $S_0$ and $S_1$ as follows.
\begin{align}
    S_0 &= \{\mathbf{a} \in \{0,1\}^n: f(\mathbf{a}) =0\},\\
    S_1 &= \{\mathbf{a} \in \{0,1\}^n: f(\mathbf{a}) =1\}.
\end{align}
The polynomial $p$ is then constructed as 
\begin{align}
    f(x_1,\ldots,x_n) & = p(x_1,\ldots,x_n, y_1,\ldots,y_n) \nonumber \\ & = 
    \sum_{\mathbf{a} \in S_1} h_{\mathbf{a}} = 1 + \sum_{\mathbf{a} \in S_0} h_{\mathbf{a}},\label{eq:boolPoly}
\end{align}
where $y_i = x_i+1$.

We note that this model applies to verifying the digital signatures, where the verification does not depend on past blocks, i.e., $P=0$. Utilizing the above technique, we can transfer any non-polynomial computations like inversions and cryptographic hash functions (e.g., SHA-2) into polynomial evaluations.

For Boolean verification polynomials over binary field as in~(\ref{eq:boolPoly}), the {\tt PolyShard} data encoding~(\ref{eq:hisEncode}) does not directly apply since it requires the underlying field size $|\mathbb{F}|$ to be at least the network size $N$. To use {\tt PolyShard} in this case, we can embed each element $y_{k}[i] \in \{0,1\}$ of a verified block $Y_k$ (time index omitted) into a binary extension field $\mathbb{F}_{2^m}$ with $2^m \geq N$. Specifically, the embedding $\bar{y}_k[i] \in \mathbb{F}_{2^m}$ of the element $y_k[i]$ is generated such that
\begin{align}\label{eq:extension}
\bar{y}_k[i]= \begin{cases} \underbrace{00\cdots0}_{m}, & y_k[i]=0,\\
\underbrace{00\cdots0}_{m-1}1, & y_k[i]=1.
\end{cases}
\end{align}
Then we can select distinct elements $\alpha_1, \ldots, \alpha_N \in \mathbb{F}_{2^m}$ to apply the encoding strategy in (\ref{eq:hisEncode}) on the block elements in the extension field. 

Verification over extension field still generates the correct result. To see that, we can easily verify that the value of the verification polynomial $p$ as constructed in (\ref{eq:boolPoly}) is invariant with the embedding operation in (\ref{eq:extension}). That is, since the polynomial $p$ is the summation of monomials in $\mathbb{F}_2$,  when we replace each input bit with its embedding, the value of $p$ equals $\underbrace{00\cdots0}_{m}$ if the verification result is $0$, and equals $\underbrace{00\cdots0}_{m-1}1$ if the result is $1$.
\section{Iterative Polyshard}\label{sec:iterative}
In Section \ref{sec:general}, we show that the number of shards can be supported by {\tt Polyshard} is up to $K_{\textup{{\tt PolyShard}}} = \lfloor \frac{(1-2\mu)N-1}{d}+1 \rfloor$, which decreases as the degree of polynomial increases. To remove this limitation, we propose iterative {\tt Polyshard} which can work with polynomials whose degree is high. The main idea of iterative {\tt Polyshard} is to represent the verification function as a {\em low-depth} arithmetic circuit which can be implemented iteratively by computing low-degree polynomials, and then apply the {\tt Polyshard} scheme to the low-degree polynomial of each iteration. 
\subsection{Arithmetic circuit modeling for verification functions}\label{subsec:arithmetic}
Before applying iterative {\tt Polyshard} to the verification functions, we model functions to some arithmetic circuits. An arithmetic circuit is a directed acyclic graph that can be used to compute a polynomial of inputs over certain field. Nodes of the graph are referred to as gates. Every node with zero indegree is an \emph{input gate} and is labeled by either a variable or an element of the underlying field. Every other node is either a \emph{addition gate} or a \emph{multiplication gate} labeled by $+$ and $\times$, respectively. Gate $u$ is a child of gate $v$ if there is a directed edge from $v$ to $u$ in the graph. Each addition (multiplication) gate computes the sum (product) of the polynomials computed by their parent gates. See Figure \ref{fig:arith_ex} for an example of an arithmetic circuit that computes the polynomial $f(x_1,x_2) = x_1^2x_2+x_1x_2$. 
To model verification functions, we consider the class of arithmetic circuits in which each layer satisfies the following conditions:
\begin{enumerate}[leftmargin=*]
\item Each multiplication gate has two inputs. 
\item Each addition gate has arbitrary number of inputs.
\item Each layer consists of addition gates followed by multiplication gates.
\item Edges within each layer are only from addition gates to multiplication gates.
\item The outputs of multiplication gates are the inputs of addition gates in the following layer.
\end{enumerate}
\begin{remark}
For any polynomial with degree $d$, there exists an arithmetic circuit satisfying the above conditions with $\lceil \log{d} \rceil+ 1$ layers. It implies that the arithmetic circuits of verification function can be low-depth. 
\end{remark}
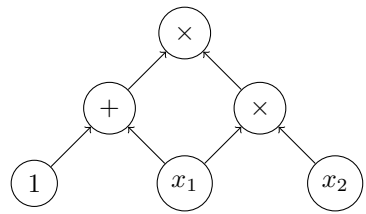
\begin{figure}[t]
\begin{center}
\begin{tikzpicture}
\tikzstyle{every node}=[draw,shape=circle];
\path (0,0) node[draw,shape=circle] (v0) {$1$};
\path (2,0) node[draw,shape=circle] (v1) {$x_1$};
\path (4,0) node[draw,shape=circle] (v2) {$x_2$};

\path (1,1) node[draw,shape=circle] (v5) {$+$};
\path (3,1) node[draw,shape=circle] (v6) {$\times$};

\draw[->] (v0) -- (v5);
\draw[->] (v1) -- (v5);
\draw[->] (v1) -- (v6);
\draw[->] (v2) -- (v6);

\path (2,2) node[draw,shape=circle] (v9) {$\times$};

\draw[->] (v5) -- (v9);
\draw[->] (v6) -- (v9);

\end{tikzpicture}
\end{center}
\caption{An arithmetic circuit that computes the polynomial $f = x_1^2x_2+x_1x_2$.}
\label{fig:arith_ex}
\end{figure}

For the arithmetic function of verification function $f^t$, we define the following terms. For each layer $l \in [1:L]$ ($L$ is the number of layers), we denote the number of multiplication gates by $A_l$. In layer $l$, there are $A_l$ intermediate polynomials (outputs of multiplication gates) denoted by $f^t_{(l,1)},\ldots,f^t_{(l,A_l)}$. The inputs of layer $l$ are denoted by $(X^l_1(t),\ldots,X^l_K(t))$. For layer $1$, we have $X^1_k(t) = \{X_k(t),Y^{t-1}_k\}$. Because of the structure of arithmetic circuits we consider, for each layer $l \in [2:L]$, we have $X^l_k(t) = \{f^t_{(l-1,1)}(X^{l-1}_k(t)),\ldots,f^t_{(l-1,A_{l-1})}(X^{l-1}_k(t))\}$.
Then, the outputs of multiplication gates in layer $L$ is the computation of verification function, i.e.,
$f^t(X_k(t),Y_k^{t-1}) = (f^t_{(L,1)}(X^l_k(t)),\ldots,f^t_{(L,A_L)}(X^l_k(t)))$.

Let's illustrate the arithmetic circuits through the following example.

\begin{figure}
\begin{center}
\begin{tikzpicture}
\path (0,0) node[draw,shape=circle] (v0) {$x_1$};
\path (2,0) node[draw,shape=circle] (v1) {$x_2$};
\path (4,0) node[draw,shape=circle] (v2) {$x_3$};
\path (6,0) node[draw,shape=circle] (v3) {$x_4$};
\path (8,0) node[draw,shape=circle] (v4) {$x_5$};

\path (1,1) node[draw,shape=circle] (v5) {$+$};
\path (3,1) node[draw,shape=circle] (v6) {$+$};
\path (5,1) node[draw,shape=circle] (v7) {$+$};
\path (7,1) node[draw,shape=circle] (v8) {$+$};

\draw[->] (v0) -- (v5);
\draw[->] (v1) -- (v5);
\draw[->] (v1) -- (v6);
\draw[->] (v2) -- (v6);
\draw[->] (v2) -- (v7);
\draw[->] (v3) -- (v7);
\draw[->] (v3) -- (v8);
\draw[->] (v4) -- (v8);

\path (2,2) node[draw,shape=circle] (v9) {$\times$};
\path (4,2) node[draw,shape=circle] (v10) {$\times$};
\path (6,2) node[draw,shape=circle] (v11) {$\times$};

\path (2.8,2) node {$f^t_{(1,1)}$};
\path (4.8,2) node {$f^t_{(1,2)}$};
\path (6.8,2) node {$f^t_{(1,3)}$};

\draw[->] (v5) -- (v9);
\draw[->] (v6) -- (v9);
\draw[->] (v6) -- (v10);
\draw[->] (v7) -- (v10);
\draw[->] (v7) -- (v11);
\draw[->] (v8) -- (v11);

\path (3,3) node[draw,shape=circle] (v12) {$+$};
\path (5,3) node[draw,shape=circle] (v13) {$+$};

\draw[->] (v9) -- (v12);
\draw[->] (v10) -- (v12);
\draw[->] (v10) -- (v13);
\draw[->] (v11) -- (v13);

\path (4,4) node[draw,shape=circle] (v14) {$\times$};

\path (4.8,4) node {$f^t_{(2,1)}$};

\draw[->] (v12) -- (v14);
\draw[->] (v13) -- (v14);

\end{tikzpicture}
\end{center}
\caption{An arithmetic circuit that computes the verification function $f^t = (x_2+x_3)\times(x_3+x_4)\times(x_1+x_2+x_3+x_4)\times(x_2+x_3+x_4+x_5)$.}
\label{fig:iter-ex}
\vspace{-3mm}
\end{figure}
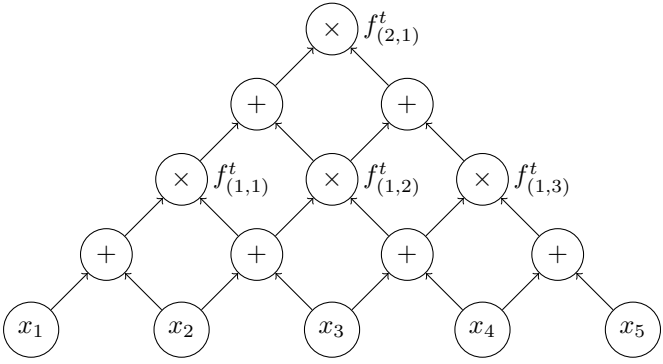

\noindent{\bf Example.} We consider verification function 
\begin{align*}
    f^t(x_1, \ldots, x_5) = (x_2+x_3)& \times(x_3+x_4)\times(x_1+x_2+x_3+x_4)\\
    & \times (x_2+x_3+x_4+x_5).
\end{align*}
An arithmetic circuit satisfying the above conditions that computes $f^t$ is depicted in Figure \ref{fig:iter-ex}. The arithmetic circuit for function $f^t$ consists two layers. The input of the first layer is $(X^1_1(t), \ldots, X^1_K(t))$, where $X^1_k(t) = X_k = (x_{k1}, \ldots, x_{k5})$, for each $k \in [1:K]$.
At the end of the first layer, $f^t_{(1,1)}(X_k), f^t_{(1,2)}(X_k), f^t_{(1,3)}(X_k)$ for each $k \in [1:K]$ are computed.
The inputs of layer $2$ is $(X^2_1(t), \ldots, X^2_K(t))$, where $X^2_k(t) = (f^t_{(1,1)}(X_k), f^t_{(1,2)}(X_k), f^t_{(1,3)}(X_k))$.
At the end of the layer $2$, $f^t_{(2,1)}(X^2_k(t)) = f^t_{(2,1)}(f^t_{(1,1)}(X_k), f^t_{(1,2)}(X_k), f^t_{(1,3)}(X_k)) = f(X_k)$, for each $k \in [1:K]$ are computed.

\subsection{Iterative coded verification} \label{subsec:iterative}
The block verification process of iterative {\tt PolyShard} has $L$ iterations. Each iteration $l\in[1:L]$ has the following three steps.

\noindent {\bf Step 1: block encoding.} Each node~$i$ generates a coded block $\tilde{X}^l_i(t) = v^l_t(\alpha_i)$ where $v^l_t(z) = \sum_{k=1}^K X^l_k(t) \prod_{j \neq k} \frac{z - \omega_j}{\omega_k - \omega_j}$ as a linear combination using the same set of coefficients in (\ref{eq:hisEncode}), i.e., $\tilde{X}^l_i(t) = v^l_t(\alpha_i) = \sum_{k=1}^K \ell_{ik} X^l_k(t)$. This step incurs $O(NK)$ operations across the network, since each of the $N$ nodes computes a linear combination of $K$ blocks.

\noindent {\bf Step 2: coded computation.} Each node $i$ applies intermediate functions $f^t_{(l,1)},\ldots,f^t_{(l,A_l)}$ directly to the coded input $\tilde{X}^l_i(t)$ to compute $g^t_{(l,i)}= (f^t_{(l,1)}(\tilde{X}^l_i(t)),\ldots,f^t_{(l,A_l)}(\tilde{X}^l_i(t)))$, and broadcasts it to all other nodes. We note that the total computational complexity of one node incurred in this step is the complexity of computing arithmetic circuit of $f^t$. We assume that the computational complexity of arithmetic circuit of function $f^t$ $\approx c(f^t)$.

\noindent {\bf Step 3: decoding.} Since each intermediate function $f^t_{(l,a)}$ is a polynomial of degree $2$ over the inputs of layer $l$. To decode $f^t_{(l,1)},\ldots,f^t_{(l,A_l)}$ from $g^t_{(l,1)},\ldots,g^t_{(l,N)}$, each node needs to decode a Reed-Solomon code with dimension $2(K-1)+1$ and length $N$. To successfully decode the results, it requires the number of errors $\mu N \leq (N-2(K-1)-1)/2$. That is, the maximum number of shards $K$ that can be securely supported is $K_{\textup{{\tt Iterative}}} = \lfloor \frac{(1-2\mu)N+1}{2} \rfloor$. 

After computation of $L$ iterations, each node evaluates $f^t_{(L,1)}(v^l_t(z)),\ldots,f^t_{(L,A_L)}(v^l_t(z))$ at $\omega_1,\ldots,\omega_K$ to recover $\{(f^t_{(L,1)}(X^l_k(t)),\ldots,f^t_{(L,A_L)}(X^l_k(t)))\}^K_{k=1}=\{f^t(X_k(t),Y_k^{t-1})\}_{k=1}^K$, to obtain the verification results $\{e_k^t\}_{k=1}^K$ and the verified blocks $\{Y_k(t) = e_k^t X_k(t)\}_{k=1}^K$. Finally, each node $i$ computes $\tilde{Y}_i(t)$ following (\ref{eq:hisEncode}), and appends it to its local coded sub-chain which incurs the computational complexity which is $O(NK)$.

\subsection{Performance of iterative {\tt Polyshard}}
As shown in Theorem \ref{thm:IterPolyShard}, we have that the maximum number of shards supported by iterative {\tt Polyshard} is independent of the degree of verification function. 
Moreover, iterative {\tt PolyShard} achieves a storage efficiency $\gamma_{\textup{{\tt Iterative}}} = \gamma_{\textup{{\tt Polyshard}}} = \Theta(N)$, and it is also robust against $\mu N = \Theta(N)$ quickly-adaptive adversaries. Then, the total number of operations during the verification and the sub-chain update processes is $O(NK) + Nc(f^t) + O(N^2 \log^2 N \log\log N)$. The coding overhead reduces to $O(N^2 \log^2 N \log\log N)$ since $K_{\textup{{\tt Iterative}}}\leq N$. The throughput of iterative {\tt PolyShard} is computed as
\begin{align}
    \lambda_{\textup{{\tt Iterative}}} 
&=\liminf_{t \rightarrow \infty} \frac{K_{\textup{{\tt Iterative}}}}{1 + \frac{O(N\log^2 N \log \log N)}{c(f^t)}}.
\end{align}
When $c(f^t)$ grows with $t$, the throughput of iterative ${\tt PolyShard}$ becomes $ \lambda_{\textup{{\tt Iterative}}}=\Theta(N)$, i.e., iterative {\tt PolyShard} simultaneously achieves the information-theoretically optimal storage efficiency, security, and throughput to within constant multiplicative gaps.



\end{document}